\documentclass[aps,onecolumn,showpacs, preprintnumbers,amsmath,amssymb]{revtex4}

\usepackage{graphicx}
\usepackage{amsmath}
\usepackage{mathbbol}
\usepackage{hyperref}

\renewcommand{\rho}{\varrho}
\renewcommand{\phi}{\varphi}
\renewcommand{\vec}{\mathbf}

\newcommand{\ee}{\mathbb{e}}

\begin{document}

\title{Atomic models of dense plasmas, applications and current challenges}

\author{R. Piron$^{1,2}$\footnote{Corresponding author\\Electronic address: robin.piron@cea.fr}}
\affiliation{$^1$CEA, DAM, DIF, F-91297 Arpajon, France.}
\affiliation{$^2$Universit\'{e} Paris-Saclay, CEA, Laboratoire Mati\`{e}re en Conditions Extr\^{e}mes, F-91680 Bruy\`{e}res-le-Ch\^{a}tel, France.}


\date{\today}

\begin{abstract}

Modeling plasmas in terms of atoms or ions is theoretically appealing for several reasons. When it is relevant, the notion of atom or ion in a plasma provides us with an interpretation scheme of the plasma's internal functioning. From the standpoint of quantitative estimation of plasma properties, atomic models of plasma allow to extend many theoretical tools of atomic physics to plasmas. This notably includes the statistical approaches to the detailed accounting for excited states, or the collisional-radiative modeling of non-equilibrium plasmas, which is based on the notion of atomic processes. In this paper, we focus on the challenges raised by the atomic modeling of dense, non-ideal plasmas.

First we make a brief, non-exhaustive review of atomic models of plasmas, from ideal plasmas to strongly-coupled and pressure-ionized plasmas. We discuss the limitations of these models and pinpoint some open problems in the field of atomic modeling of plasmas.

We then address the peculiarities of atomic processes in dense plasmas and point out some specific issues relative to the calculation of their cross-sections. In particular, we discuss the modeling of fluctuations, the accounting for channel mixing and collective phenomena in the photoabsorption, or the impact of pressure ionization on collisional processes.

\end{abstract}


\maketitle

\section{Introduction}
Plasma physics, in its generalty, deals with the physics of partially- or fully- ionized matter, not having periodic structure (i.e. ionized fluids). Fundamentally, a plasma is just a collection of a large number of electrons and nuclei interacting through Coulomb potentials, both repulsive and attractive. 
The notions of atom, ion, free-electron, and ionization are interpreting schemes that may emerge from particular models, and should not be seen as objects whose existence is pre-supposed. 

In the following we focus on the case of homogeneous plasmas in thermodynamic equilibrium. In this context, homogeneous means that the thermodynamic averages of the particle densities has no spatial variation. Any spatial inhomogeneity or \emph{structure} (electronic structure or fluid structure) that is considered is then a matter of \emph{correlations} among the particle positions.

Due to the existence of long-range attractive potentials in the system, a first-principle modeling of plasmas based on classical mechanics only is virtually impossible, because of the classical Coulomb catastrophe. It requires either an \textit{ad hoc} modification of the potential at short distances (regularization, cut-off, smearing of charges...)
 or a well-chosen hypothesis that allows to circumvent this problem. In a sense, whatever the thermodynamic conditions, the quantum behavior of the electrons is always essential in founding the stability of plasmas \cite{Dyson67,Lenard68,Lebowitz69,Lieb72,Lieb76}.

Among the various models that may be used to describe plasmas, models which define a notion of atom or ion are particularly appealing. We mean here a system that describes the plasma using effective 1-electron states (orbitals) stemming from a spherically-symmetric 1-electron Hamiltonian. With such kind of model, one can greatly reduce the complexity of the description of the microstates of the plasma, using the notion of atomic excited states.

A description of the atomic excited states in the plasma is essential to the modeling of spectral quantities, such as radiative properties, which can reveal the fluctuations around the average atomic state. More generally, the notion of atomic process founds the approaches to many phenomena. One may cite, for instance, the collisional-radiative modeling of plasmas out of local thermodynamic equilibrium \cite{MihalasStellarAtmospheres,Ralchenkobook}, or the approaches to line-broadening mechanisms \cite{Baranger58,Stambulchik08}.

Being spherically symmetric, the effective 1-electron Hamiltonian commutes with angular momentum operators, which enables the separation of the angular part of the 1-electron states (spherical harmonic or spinors). Many-electron states can then be built from the orbitals using a well-established mathematical apparatus, with the Wigner-Eckart theorem as a cornerstone. Methods of angular-momentum coupling where thoroughly studied, both as regards the formalism \cite{Racah42,Racah42b,Racah43,Judd,Yutsis} and the numerical methods \cite{BarShalom88,Gaigalas01}. 

Due to the tremendous number of populated atomic states at high temperatures, statistical grouping of energy levels is particularly useful for hot-plasma modeling. Analytic results from atomic physics are available for the statistical properties of configurations \cite{Moszkowski62} and transition arrays between configurations \cite{Bauche79,Bauche82,Bauche85} (UTA, SOSA). This enables powerful detailed configuration accounting (DCA) approaches. In order to handle the calculation of spectra for ions with highly complex electronic structure, coarser statistical approaches were developped, such as the gaussian approximation \cite{Perrot88b} and the super transition arrays (STA) \cite{BarShalom89,Blenski00}. On the other hand, in order to refine over the UTA approach, finer approaches were also developped, such as the partially-resolved transition arrays (PRTA) \cite{Iglesias12}.



\section{Atomic modeling of ideal plasmas\label{sec_ideal_plasmas}}

The simplest hypothesis allowing to circumvent the classical catastrophe is to neglect all interactions in the plasma (ideal-gas approximation). Applying this approximation directly to the system of nuclei and electrons yields an ideal-gas model valid for a fully-ionized plasma at zero coupling.

Usually, in plasma physics, the ideal-gas approximation is only partially applied. One first defines a quasi-particle composed of a nucleus and a set of bound electrons which interact with the nucleus and among themselves. This is the definition of an \emph{ion} in this model. All electrons of the plasma that are not bound to a nucleus are viewed as ``free'' electrons. One then makes the ideal-gas hypothesis on the system of ions and free electrons. This is the picture of an \emph{ideal plasma}. 

In this type of model, one usually resorts to a quantum model for the electronic structure of the ion, which of course does not lead to the Coulomb catastrophe. On the other hand, the ideal-gas approximation allows to circumvent the Coulomb catastrophe for the whole plasma of ions and free electrons.

In this context, the ion is seen as a charged, isolated system having a finite spatial extension, that is an \emph{isolated} ion. To model its electronic structure requires to address a $N$-body problem with $N \sim Z+1$ at most.

\subsection{Saha-Boltzmann model and isolated ion: a variational detailed model of ideal plasma}
The approach in which one accounts for the various electronic states of the plasma as different species is known as \emph{detailed} modeling (also sometimes called ``chemical approach''). For an ideal plasma, this approach yields the Saha-Boltzmann model of plasma \cite{Saha20,Saha21}.

The modeling of the isolated-ion electronic structure is treated as a problem independent from the modeling of the whole plasma. A typical quantum Hamiltonian of the isolated ion having $Q$ bound electrons is the $Q$-electron operator:
\begin{align}
\hat{H}_\text{isol.ion}^Q=\sum_{j=1}^{Q}\frac{\tilde{\vec{P}}_j^2}{2m_\text{e}}-\sum_{j=1}^{Q}\frac{Z \ee^2}{|\tilde{\vec{R}}_j|}
+\sum_{j=1}^{Q}\sum_{\substack{k=1\\k\neq j}}^{Q}\frac{\ee^2}{|\tilde{\vec{R}}_j-\tilde{\vec{R}}_k|}
\end{align}
where $\tilde{\vec{P}}_j$ is the 1-electron momentum operator acting on the $j$ electron, $\tilde{\vec{R}}_j$ is the 1-electron position operator acting on the $j$ electron.

The problem of finding the stationnary states of this isolated-ion Hamiltonian has a variational formulation (Ritz theorem). However, to tackle this problem, one  resorts to approximate methods. Such methods often starts with a model based on effective 1-electron states $\varphi_\xi$ (i.e. orbitals) which are solutions of a 1-electron Schr\"{odinger} equation associated to an effective spherically-symmetric potential $v_\text{eff}(r)$.
\begin{align}
-\frac{\hbar^2}{2m_\text{e}}\nabla_{\vec{r}}^2\varphi_\xi(\vec{r})+v_\text{eff}(r)\varphi_\xi(\vec{r})=\varepsilon_\xi \varphi_\xi(\vec{r})
\label{eq_general_eff_pot}
\end{align}
Various models exists, which mostly differ in the way of obtaining the effective potential (Hartree-Fock-Slater \cite{Slater51}, optimized effective potential \cite{Talman76}, parametric potential \cite{Klapisch71}...). 
In the Hartree-Fock model \cite{Slater28}, a nonlocal exchange term is added to Eq.~\eqref{eq_general_eff_pot}. However, the problem is often restricted to an effective 1-electron problem having spherical symmetry (restricted Hartree-Fock approach, see for instance \cite{Cowan}).

One then refine over this approximate description using perturbation theory, performing the diagonalization of the many-electron perturbation operators in a many-electron basis built from the orbitals (see, for instance \cite{FroeseFischerbook,Cowan,Grantbook}). Depending on the approach used, such approximate model may have a variational formulation or not. 

The approximate many-electron states are naturally grouped into degenerate energy levels, which in practice depend on the coupling scheme that is chosen. Accounting for each level as a species is called a detailed level accounting (DLA) approach. One may also perform a detailed configuration accounting (DCA), grouping the levels according to their parent configuration \cite{Slater29}. One may even group the configurations into broader statistical objects: superconfigurations \cite{BarShalom89}. Each degree of statistical grouping goes along with a set of ion species $\alpha$ to account for, each having as main properties a mean energy $E_\alpha$ and a degeneracy $g_\alpha$. Usually the reference of energies is taken such as $E_\alpha = 0$ for the bare nucleus and the free electrons.

The classical Hamiltonian for the ion-free-electron ideal plasma is the following:
\begin{align}
\mathcal{H}_\text{id}(\{p_{\alpha,j}\})=\sum_{\alpha=1}^{M,\text{e}}\sum_{j=1}^{N_\alpha}\left(\frac{p_{\alpha,j}}{2 m_\alpha}+E_\alpha\right)
\end{align}
where the first sum runs over the $M$ ion species plus the free-electron species labelled by``e''. In this Hamiltonian, interactions among the particles of the plasma are neglected. The ion-free-electron system is then considered in the canonical ensemble. The free energy per ion of the classical ideal-gas mixture is: 
\begin{align}
\dot{F}_\text{id}\left(\{n_\alpha\},T\right)
&=-\sum_{\alpha=1}^{M,\text{e}} \frac{n_\alpha}{n_\text{i}\beta} 
\left( \log\left(\frac{g_\alpha e^{-\beta E_\alpha}}{n_\alpha\Lambda_\alpha^3}\right) + 1 \right)
\label{eq_Saha_free_energy}
\end{align}
where $n_\alpha$ is the number of particle of species $\alpha$ per unit volume (in particular, $n_\text{e}$ is the free-electron density), $n_\text{i}=\sum_{\alpha=1}^{M}{n_\alpha}$, and $\Lambda_\alpha=h/\sqrt{2\pi m_\alpha k_\text{B} T}$ is the classical thermal length.


The equations of the Saha equilibrium model are obtained through a minimization of the free energy of the system with respect to the species populations, while also requiring the neutrality of the plasma and a fixed number of ions. Thus, transfers of population among the various ion species are allowed, and the populations are ultimately set by the condition of thermodynamic equilibrium. Only the populations, and not the quantities related to the shell structure of the ions stem from the model.
\begin{align}
\dot{F}_\text{eq}(n_\text{i},T)
=\underset{n_\alpha,n_\text{e}}{\text{Min}}\,\dot{F}\left(\{n_\alpha\},T\right)
~&\text{ s. t. } \sum_\alpha n_\alpha =n_\text{i}
\nonumber\\
&\text{ s. t. } \sum_\alpha n_\alpha Z_\alpha^* =n_\text{e}
\end{align}

The latter minimization yields the following condition on the chemical potentials:
\begin{align}
&\text{for all }\alpha\neq \text{e},\ 
\mu_{\text{id},\alpha}(n_\alpha,T)+\mu_{\text{id,e}}(n_\text{e},T) Z_\alpha^* =\lambda_\text{i}\text{ independent of $\alpha$}
\label{eq_Saha_chemical_equilibrium}
\end{align}
with the classical-ideal-gas chemical potentials being:
\begin{align}
&\mu_{\text{id},\alpha}(n_\alpha,T)
=\frac{\partial}{\partial n_\alpha}\left( n_\text{i}\dot{F}\left(\{n_\alpha\},T\right)\right)
=\frac{1}{\beta} 
\log\left(\frac{g_\alpha e^{-\beta E_\alpha} }{n_\alpha\Lambda_\alpha^3}\right)
\label{eq_id_chem_pot}
\end{align}
For a plasma of a pure substance, one usually assumes the thermal lengths $\Lambda_\alpha$ of all ion species to be equal: $\Lambda_\alpha\approx\Lambda_\text{i}$. From Eq.~\eqref{eq_Saha_chemical_equilibrium} and the two contrains, one gets the populations:
\begin{align}
&\frac{n_\alpha}{n_\text{i}}=\frac{g_\alpha e^{-\beta(E_\alpha-\mu_{\text{id,e}}Z_\alpha^*)}}
{\sum_{\gamma=1}^M g_\gamma e^{-\beta(E_\gamma-\mu_{\text{id,e}}Z_\gamma^*)}}
\label{eq_id_Saha_pops}\\
&\sum_\alpha n_\alpha Z_\alpha^* =n_\text{e}
\end{align}

From these populations follows notably the mean ionization of the plasma $Z^*=n_\text{e}/n_\text{i}$, as a value set by the thermodynamic equilibrium condition. Obtaining the ionization state of the plasma as a result of its equilibrium state is among the purpose of plasma modeling. 

The plasma being treated in the ideal-gas approximation, the pair distribution functions among the particle of the plasma are identically $g_{\alpha,\gamma}=1$.

The thermodynamic quantities stemming from this model (internal energy, pressure...) can be obtained from the free energy, using the revelant derivatives. They simply correspond to those of the ideal-gas mixture, taken with the equilibrium values of the species populations. They obviously fulfill the virial theorem (in its non-relativistic version):
\begin{align}
P_\text{thermo}=n_\text{i}^2\frac{\partial \dot{F}_\text{eq}(n_\text{i},T)}{\partial n_\text{i}}=
P_\text{virial}=\frac{n_\text{i}}{3}\left( 2 \dot{U}_\text{eq}(n_\text{i},T)-\dot{U}_\text{inter,eq}(n_\text{i},T) \right)
\label{eq_virial_theorem}
\end{align}
where $\dot{U}_\text{eq}$ denotes the internal energy per ion and $\dot{U}_\text{inter,eq}$ denotes the interaction energy per ion, which is zero for the ideal-gas mixture. This is an important feature as regards the consistency of the equation of state. 

An issue with the present model is that, in principle, when accounting for the excited states in a complete manner, the partition function diverges because of the infinite number of states. As noted by Urey \cite{Urey24}, Fermi \cite{Fermi24} and Larkin \cite{Larkin60}, the solution to this puzzle is precisely to be found in the non-ideal corrections to the Saha equilibrium. These restrict somehow the set of states to be accounted for in the calculation. We will elaborate more on this point at the end of the following section.

Finally, let us mention the straightforward extension of the Saha model that consists in replacing the classical-ideal-gas free energy for the free electrons by that of the Fermi ideal gas.
\begin{align}
&\dot{F}\left(\{n_\alpha\},T\right)
=-\sum_{\alpha=1}^{M} \frac{n_\alpha}{n_\text{i}\beta} 
\left( \log\left(\frac{g_\alpha e^{-\beta E_\alpha}}{n_\alpha\Lambda_\alpha^3}\right) + 1 \right)
+\frac{f_\text{e}^\text{F}(n_\text{e},T)}{n_\text{i}}
\label{eq_Saha_Fermi_gas}\\
&f_\text{e}^\text{F}(n_\text{e},T)=n_\text{e}\mu^\text{F}_\text{e}(n_\text{e},T)-\frac{2}{3}u^\text{F}_\text{e}(n_\text{e},T)\\
&u^\text{F}_\text{e}(n_\text{e},T)=\frac{4}{\beta\sqrt{\pi}\Lambda_\text{e}^3}I_{3/2}\left(\beta\mu^\text{F}_\text{e}(n_\text{e})
\right)\\
&n_\text{e}=\frac{4}{\sqrt{\pi}\Lambda_\text{e}^3}I_{1/2}\left(\beta\mu^\text{F}_\text{e}(n_\text{e},T)\right)
\end{align}
where the sum in Eq.~\eqref{eq_Saha_Fermi_gas} only runs over the $M$ ion species. $f_\text{e}^\text{F}(n_\text{e},T)$ and $u^\text{F}_\text{e}(n_\text{e},T)$ are the free and internal energies per unit volume of a Fermi gas of density $n_\text{e}$ and temperature $T$, respectively. $\mu^\text{F}_\text{e}(n_\text{e},T)$ is the canonical chemical potential.


\subsection{Average-atom model of isolated ion from a variational perspective\label{sec_AAII}}
The case of an ideal plasma of isolated ions can also be addressed through an \emph{average-atom} approach. In this kind of approach, instead of accounting for the many-electron states in a detailed fashion, one only aims at describing the average many-electron state of the plasma, associating fractional occupation numbers to the orbitals. The finite-temperature density-functional theory \cite{Hohenberg64, KohnSham65a, Mermin65} offers a sound theoretical basis for such models. In order to model an average isolated ion, we just have to restrict interactions to the ion nucleus and bound electrons, and to consider that any continuum electron participate in a uniform, noninteracting electron density $n_\text{e}$.

The free energy per ion of such a system can be written as follows:
\begin{align}
\dot{F}&\left\{\left\{p_\xi\right\},\underline{v}_\text{trial},n_\text{e};n_\text{i},T\right\}
=\dot{F}_\text{id,i}(n_\text{i},T)+\dot{F}_\text{id,e}(n_\text{e};n_\text{i},T)+\Delta F_1\left\{\left\{p_\xi\right\},\underline{v}_\text{trial}\right\}
\label{eq_AAII_free_energy}
\end{align}
where the functional dependencies are underlined. Here, the $\dot{F}_\text{id,i}$ term corresponds to the contribution of the nuclei ideal-gas and $\dot{F}_\text{id,e}$ corresponds to the contribution of the free-electron ideal gas:
\begin{align}
\dot{F}_\text{id,i}=\frac{1}{\beta}\left(\ln\left(n_\text{i}\Lambda_\text{i}^3\right)-1\right)
\ ; \ \dot{F}_\text{id,e}=\frac{n_\text{e}}{n_\text{i}\beta}\left(\ln\left(n_\text{e}\Lambda_\text{e}^3\right)-1\right)
\end{align}
where $\Lambda_\text{i}$ and $\Lambda_\text{e}$ are the nucleus and electron thermal lengths, respectively. As in the Saha model, the classical ideal-gas free energy of the electrons may also be replaced by the Fermi ideal-gas expression:
\begin{align}
\dot{F}_\text{id,e}=\frac{f_\text{e}^\text{F}(n_\text{e},T)}{n_\text{i}}
\end{align}

The $\Delta F_1$ term corresponds to the free-energy of the average ion electronic structure, that is: the interacting system of bound electrons and the nucleus. We treat this system using the Kohn-Sham method \cite{KohnSham65a}, that is, we split $\Delta F_1$ into the three contributions:
\begin{align}
\Delta F_1\left\{\left\{p_\xi\right\}, \underline{v}_\text{trial};T\right\}
=\Delta F_1^0 + \Delta F_1^\text{el} + \Delta F_1^\text{xc}
\end{align}
where $\Delta F_1^0$ is the kinetic-entropic contribution to the free energy of a system of independent bound electrons, feeling an external potential $\underline{v}_\text{trial}(r)$ and having 1-electron orbital populations $\{p_\xi\}$ that together yield the density $n(r)$ of the interacting bound electrons. $\Delta F_1^\text{el}$ is the direct electrostatic contribution, and $\Delta F_1^\text{xc}$ is the contribution of exchange and correlation to the free energy of the electronic structure. 

According to its definition, the expression of $\Delta F_1^0$ is:
\begin{align}
\Delta F_1^0\left\{\left\{p_\xi\right\}, \underline{v}_\text{trial};T\right\}
&=\sum_\xi \left(
p_\xi\left(\varepsilon_\xi-\int d^3r \left\{v_\text{trial}(r)|\varphi_\xi(\vec{r})|^2\right\}\right)
-T s_\xi\right)\\
&=\sum_\xi \left(
p_\xi\varepsilon_\xi-\int d^3r \left\{n(r)v_\text{trial}(r)\right\}
-T s_\xi\right)\label{eq_AAII_dF10}
\end{align}
where the sum over the $\xi$-indices only runs over the bound 1-electron states (bound orbitals). $p_\xi$ is the mean occupation number of the 1-electron state $\xi$, and the corresponding contribution to the entropy of the effective non-interacting system is:
\begin{align}
s_\xi=s(p_\xi)=-k_\text{B}\left(p_\xi\ln\left(p_\xi\right)
+(1-p_\xi)\ln\left(1-p_\xi\right)\right)
\label{eq_dft_entropy}
\end{align}
$\varepsilon_\xi$ and $\varphi_\xi(\vec{r})$ are shorthand notations for $\varepsilon_\xi\left\{\underline{v}_\text{trial}\right\}$ and $\varphi_\xi\left\{\underline{v}_\text{trial};\vec{r}\right\}$, respectively. These are the eigenvalues and wave-functions of the 1-electron states obtained in the trial potential $v_\text{trial}(r)$. In the non-relativistic case, they are obtained solving the 1-electron Schr\"{o}dinger equation:
\begin{align}
-\frac{\hbar^2}{2m_\text{e}}\nabla_{\vec{r}}^2\varphi_\xi(\vec{r})+v_\text{trial}(r)\varphi_\xi(\vec{r})=\varepsilon_\xi \varphi_\xi(\vec{r})
\end{align}
We take the convention of normalizing the $\varphi_\xi$ to unity.
The trial potential $v_\text{trial}(r)$ and occupation numbers $\{p_\xi\}$ are such that the density of the system of independent bound electrons is $n(r)$. In this context, $n(r)$ is a shorthand notation for $n\left\{\left\{p_\xi\right\},\underline{v}_\text{trial};r\right\}$:
\begin{align}
n\left\{\underline{v}_\text{trial},\{p_\xi\};r\right\}=\sum_\xi p_\xi|\varphi_\xi(\vec{r})|^2\label{eq_AAII_n}
\end{align}

The direct electrostatic contribution $\Delta F_1^\text{el}$ can be written as a functional of $n(r)$:
\begin{align}
\Delta F_1^\text{el}&\left\{\left\{p_\xi\right\}, \underline{v}_\text{trial}\right\}
=\tilde{\Delta F}_1^\text{el}\left\{\underline{n}\right\}
=\ee^2\int d^3r\left\{\frac{-Z n(r)}{r}\right\}
+\frac{\ee^2}{2}\int d^3r d^3r'\left\{\frac{n(r)n(r')}{|\vec{r}-\vec{r}'|}\right\}
\label{eq_AAII_dF1el}
\end{align}
The exchange-correlation contribution can be approximated by:
\begin{align}
\Delta F_1^\text{xc}\left\{\left\{p_\xi\right\}, \underline{v}_\text{trial};T\right\}
&=\tilde{\Delta F}_1^\text{xc}\left\{\underline{n};T\right\}
=\int d^3r\left\{f_\text{xc}\left(n(r),T\right)\right\}\label{eq_AAII_dF1xc}
\end{align}
where $f_\text{xc}$ is the exchange-correlation free energy per unit volume of an homogeneous electron gas (local density approximation). 

We stress that in this model, as in the Saha model, there is a strong distinction between bound and free electrons, since any electron that belongs to the continuum is considered as non-interacting, whereas bound electrons participate in $n(r)$ and interact both with the other bound electrons of the same ion and with its nucleus, as is seen from the expressions of $\Delta F_1^\text{el}$ and $\Delta F_1^\text{xc}$.

In order to obtain the equations of the model, we minimize the free energy per ion, requiring the additional constrain of overall neutrality:
\begin{align}
\dot{F}_\text{eq}(n_\text{i},T)
=\underset{p_\xi,v_\text{trial},n_\text{e}}{\text{Min}}\,&\dot{F}\left\{\{p_\xi\},\underline{v}_\text{trial},n_\text{e};n_\text{i},T\right\}
~\text{ s. t. } Z - \sum_\xi p_\xi =\frac{n_\text{e}}{n_\text{i}}
\label{eq_minimization_AAII}
\end{align}
Performing this constrained minimization, we get the equations of the average-atom model of isolated-ion (AAII):
\begin{align}
&v_\text{trial}(r)=v_\text{el}(r)+\mu_\text{xc}\left(n(r),T\right)\label{eq_AAII_vtrial}\\
&p_\xi =p_\text{F}(\mu,T,\varepsilon_\xi) = \frac{1}{e^{\beta(\varepsilon_\xi-\mu)}+1}\label{eq_AAII_FermiDirac}\\
&\mu=\mu_\text{id,e}(n_\text{e},T)&\text{ (classical ideal gas)}
\label{eq_AAII_mu}
\\
&\hphantom{\mu}=\mu_\text{id,e}^\text{F}(n_\text{e},T)&\text{ (Fermi ideal gas)}
\label{eq_AAII_mu_Fermi}
\end{align}
where $\mu_\text{xc}(n,T)=\partial f_\text{xc}(n,T)/\partial n$, and where $v_\text{el}(r)$ is a shorthand notation for $v_\text{el}\left\{\underline{n};r\right\}$, defined as follows:
\begin{align}
v_\text{el}\left\{\underline{n};r\right\}
=\frac{\delta \tilde{\Delta F}_1^\text{el}}{\delta n(r)}=-\frac{Z\ee^2}{r}+\ee^2\int d^3r'\left\{\frac{n(r')}{|\vec{r}-\vec{r}'|}\right\}
\label{eq_def_vel}
\end{align}

From the free energy at equilibrium, we can derive all the thermodynamic quantities of interest, in particular the pressure:
\begin{align}
P=n_\text{i}^2\frac{\partial \dot{F}_\text{eq}}{\partial n_\text{i}}
&=n_\text{i}k_\text{B} T+n_\text{e}k_\text{B} T &\text{(classical ideal gas)}
\label{eq_AAII_pressure_classical}
\\
&=n_\text{i}k_\text{B} T
-f_\text{e}^\text{F}(n_\text{e},T)
+n_\text{e}\mu^\text{F}_\text{e}(n_\text{e},T) &\text{(Fermi ideal gas)}
\label{eq_AAII_pressure_Fermi}
\end{align}
This corresponds to the pressure of the ideal-gas mixture of ions and free-electrons. It may be shown easily (method described in \cite{Slater33}) that the virial theorem is fulfilled in this model.

To conclude about this derivation, let us note that instead of considering as variables the arbitrary populations ${p_\xi}$ and trial potential $v_\text{trial}(r)$, one can consider the electron density $n(r)$ as the variable, formally inverting the relation between $v_\text{trial}(r)$ and $n(r)$. In this case, one define $v_\text{trial}\{\underline{n},n_\text{e};r\}$ as the external potential yielding the density $n(r)$ for a system of independent particle at equilibrium, that is with $p_\xi=p_\text{F}(\mu,T,\varepsilon_\xi)$. This corresponds more closely to the usual standpoint of density functional theory and we will use this one in the following derivations.

\begin{figure}[t]
\centerline{\includegraphics[width=8cm]{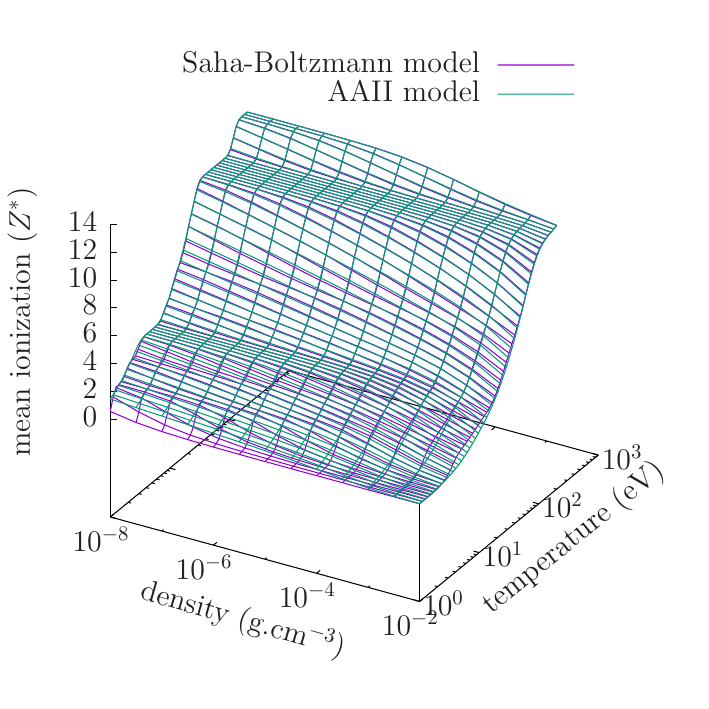}}
\caption{Mean ionization of silicon stemming from the Saha equilibrium model, with detailed configuration accounting of the ion electron states, and from the average-atom model of isolated ion (AAII).
\label{fig_Saha_AAII}}
\end{figure}

Fig.~\ref{fig_Saha_AAII} presents a comparison between the mean ionization: $Z^*\equiv n_\text{e}/n_\text{i}$ obtained from the Saha equilibrium model, using a detailed configuration accounting for the ion electron states, and the present AAII model, for the case of silicon. Though the results differ slightly, they are rather close. The qualitative behavior of decreasing mean ionization when density increases is similar, exhibiting clearly the lack of pressure ionization in these models.




\section{Nonideality corrections to isolated ions\label{sec_nonideality_corrections}}

\subsection{General considerations, notion of ionization-potential depression}
Nonideal plasmas specifically correspond to those plasmas for which the interactions of the ions with the surrounding ions and free electrons cannot be disregarded.
In first approximation, one may assume that the internal structure of the ions remains untouched and simply refine over the ideal-gas approximation for the system of point-like ions and free electrons (ion-free-electron plasma), by accounting for the interaction energy. The classical Hamiltonian for this interacting ion-free-electron plasma is the following:
\begin{align}
&\mathcal{H}(\{\vec{p}_{\alpha,j},\vec{r}_{\alpha,j}\})=
\mathcal{H}_\text{id}(\{\vec{p}_{\alpha,j}\})
+W_\text{IFE}(\{\vec{r}_{\alpha,j}\})\\
&W_\text{IFE}(\{\vec{r}_{\alpha,j}\})
=\frac{\ee^2}{2}
\sum_{\alpha=1}^{M,\text{e}}\sum_{j=1}^{N_\alpha}
\mathop{ \sum_{\gamma=1}^{M,\text{e}}\sum_{k=1}^{N_\gamma} }_{(\gamma,k)\neq (\alpha,j)}
\frac{Z^*_\alpha Z^*_\gamma}{| \vec{r}_{\alpha,j} - \vec{r}_{\gamma,k} |}
\end{align}
where the charge of a free electron is $Z^*_\text{e}=-1$.

The ions being assumed point-like, this approach to interactions can only be sensitive to the ion charge, rather than to the detail of its electronic structure. All species $\alpha$ sharing a same ion charge $Z^*_\alpha$ behave the same. Consequently, effective changes only occur in energy differences among different charge states, yielding the notion of ionization-potential depression (IPD) or continuum lowering.

Two ways of accounting for this interaction energy are often described in the literature. These two mostly lead to the same results, and rather pertain to different standpoints on the problem than to strictly distinct approaches.

A first approach is to evaluate the average potentials $v_\alpha^*$ acting on the various point-like ions of the plasma, due to the interactions with the all the other plasma particles. Typically, if we set $v_\alpha(\vec{r})$ to be the average potential around a particle of species $\alpha$, we have:
\begin{align}
v_\alpha^*=\lim_{r\rightarrow 0} \left( v_\alpha(\vec{r})-\frac{Z_\alpha \ee^2}{r}\right)
\end{align}
 
For each ion, one then adds the corresponding $v_\alpha^*$ potential to the binding energy $E_\alpha$ found from the Hamiltonian $\hat{H}_\text{isol.ion}^Q$ pertaining to the electronic structure. This may be interpreted as adding a constant perturbing potential to the Hamiltonian. This interpretation is for instance used in \cite{StewartPyatt66}. Consistently with the point-like ion hypothesis, the nucleus and each bound electron of the ion feels the same perturbing potential.  In the detailed approach, this results in substituting for the energies $E_\alpha$ in Eq.~\eqref{eq_id_Saha_pops}:
\begin{align}
E_\alpha^*=E_\alpha+Z^*_\alpha v_\alpha^*
\end{align}
The correction to the ionization potential is thus:
\begin{align}
\Delta I_\alpha&=Z^*_\gamma v_\gamma^* - Z^*_\alpha v_\alpha^* \text{ with } Z^*_\gamma=Z^*_\alpha+1
\\
&\approx v_\gamma^*
\end{align}
In the average-atom approach, this amounts to substituting for the energies of the orbitals, in Eq.~\eqref{eq_AAII_FermiDirac}:
\begin{align}
\varepsilon_\alpha^*=\varepsilon_\alpha+v^*
\end{align}
with $v^*$ calculated for the average ion charge $Z^*$.
 
However, this does not correspond to a unified treatment of both the atomic structure of the ions and the interactions among the particles of the plasma. Such a transposition of the interaction potentials in the plasma, stemming from a particular model, to the effective potentials pertaining to the ions electronic structure is rather heuristical than formally justified. Moreover, this standpoint on the treatment of interactions in the plasma can lead to some confusion, as it may appear as accouting for the effect of interactions on the ion electronic structure, which it does not really do. This procedure shifts the energy reference of each charge state, without modifying the spectrum or the orbitals of the electronic structure \textit{per se}.

A slightly different standpoint, maybe less heuristical, on the accounting for the interaction energy is to add an approximate interaction contribution to the free energy of the ideal plasma. In statistical mechanics, such contribution is called an excess free energy.

In the detailed approach, Eq.~\eqref{eq_Saha_free_energy} becomes:
\begin{align}
&F\left(\{n_\alpha\},T\right)
=F_\text{id}\left(\{n_\alpha\},T\right) + F_\text{ex}\left(\{n_\alpha\},T\right)
\label{eq_Saha_correction_Fex}
\end{align}
This results in adding excess chemical potentials $\mu_{\text{ex},\alpha}$ in the chemical-equilibrium equation Eq.~\eqref{eq_Saha_chemical_equilibrium}. This standpoint is, for instance, adopted in \cite{Griem62}.
\begin{align}
&\text{for all }\alpha\neq \text{e},~
\mu_{\text{id},\alpha}(n_\alpha,T)+\mu_{\text{ex},\alpha}(\{n_\alpha\},T)
+\left(\mu_{\text{id,e}}(n_\text{e},T)+\mu_{\text{ex,e}}(\{n_\alpha\},T)\right) Z_\alpha^* =\lambda_\text{i}
\label{eq_Saha_chemical_equilibrium_excess}
\end{align}
In this framework, the effective correction to the ionization potential appears when one calculate, for instance the ratio of populations for levels belonging to neighbouring charge states. It results from the corrections to the chemical potentials of the neighbouring charge states, and free electrons:
\begin{align}
\Delta I_\alpha
&= \mu_{\text{ex},\gamma}-\mu_{\text{ex},\alpha}+\mu_{\text{ex},\text{e}} \text{ with $Z_\gamma=Z_\alpha+1$}
\end{align}

In the average-atom case, one simply adds the excess free energy $F_\text{ex}(n_\text{e},n_\text{i},Z^*,T)$ corresponding to the plasma of free electrons and a sole species of ions having the average charge $Z^*$, which depends on $n_\text{e}$. Eq.~\eqref{eq_AAII_free_energy} becomes:
\begin{align}
F&\left\{\left\{f_\alpha\right\},\underline{v}_\text{trial},n_\text{e};n_\text{i},T\right\}
=F_\text{id,i}(n_\text{i},T)+F_\text{id,e}(n_\text{e};n_\text{i},T)+F_\text{ex}(n_\text{e},n_\text{i},Z^*=n_\text{e}/n_\text{i},T)
+\Delta F_1\left\{\left\{f_\alpha\right\},\underline{v}_\text{trial}\right\}
\label{eq_AAII_Fex_correction}
\end{align}
and Eq.~\eqref{eq_AAII_mu} becomes
\begin{align}
\mu=\mu_\text{id,e}(n_\text{e},T)+\frac{\partial}{\partial n_\text{e}}F_\text{ex}(n_\text{e},n_\text{i},Z^*=n_\text{e}/n_\text{i},T)
\end{align}
where one must accounts for the dependency of $Z^*$ on $n_\text{e}$ when calculating the derivative.

The underlying hypothesis common to these approaches is that the ion is point-like when compared to the typical interparticle distance $d$ in the ion-free-electron plasma. 
\begin{align}
r_\text{outer} << d
\end{align}
In a detailed model, $r_\text{outer}$ is the largest mean radius of any populated orbital of any significantly-populated multi-electron state. In an average-atom model, $r_\text{outer}$ is the mean radius of the outer significantly-populated orbital. The relevant interparticle distance $d$ depends on the model used for the ion-free-electron plasma.

The point-like-ion hypothesis is the key argument in separating the modeling of interactions in the plasma from the modeling of the ion electronic structure. However, it excludes the case where some populated orbitals of an ion are perturbed by the effect of its surrounding particles, that is the case of pressure-ionized plasmas.


\subsection{Mean-field approach}
Calculating the interactions in the ion-free-electron plasma requires to address its spatial correlations functions. In classical mechanics, this may be done using the Percus picture, which relates the two-particle correlation functions in a homogeneous system to 1-particle densities in inhomogeneous fictitious systems \cite{Percus64}. 
In brief, the $\alpha,\gamma$ pair distribution function of the homogeneous fluid is related to the average density $n_{\alpha,\gamma}(r)$ of particles of species $\gamma$, around a particle of species $\alpha$ placed at the origin:
\begin{align}
&n_{\alpha,\gamma}(r)=n_\gamma g_{\alpha,\gamma}(r)\label{eq_NLDH_Percus}
\end{align}
One is then left with the problem of calculating the particle densities $n_{\alpha,\gamma}(r)$ in a set of fictitious systems, each rendered inhomogeneous by a particle of species $\alpha$ placed at the origin. The mean field (or Hartree) approach to this problem yields the nonlinear Debye-H\"{u}ckel model (NLDH model) of the ion-free-electron plasma, which can also be called Poisson-Boltzmann model of ion-free-electron plasma. The equations of this model may be summarized as follows:
\begin{align}
&\nabla^2 v_{\alpha}(r)=-4\pi \ee^2\sum_{\gamma=1}^{M,\text{e}}Z_\gamma^*  n_{\alpha,\gamma}(r) \label{eq_NLDH_Poisson}\\
&\lim_{r\rightarrow 0}v_{\alpha}(r)=\frac{Z_\alpha^* \ee^2}{r} \label{eq_NLDH_boundary}\\
&\lim_{r\rightarrow \infty}v_{\alpha}(r)=0 \label{eq_NLDH_boundary_inf}\\
&n_{\alpha,\gamma}(r)=n_\gamma e^{-\beta Z_\gamma^* v_{\alpha}(r)}
\label{eq_NLDH_Boltzmann}\\
&\sum_{\gamma=1}^{M,\text{e}}n_\gamma Z_\gamma^*=0
\label{eq_NLDH_neutrality}
\end{align}
Here, $v_\alpha(r)$ corresponds to the self-consistent-field potential around the central particle of species $\alpha$, in the related Percus fictitious system. Eq.~\eqref{eq_NLDH_Poisson} is the Poisson equation for $v_\alpha(r)$, Eqs~\eqref{eq_NLDH_boundary} and \eqref{eq_NLDH_boundary_inf} are the associated boundary conditions, Eq.~\eqref{eq_NLDH_Boltzmann} follows from the 1-particle Boltzmann distribution, and Eq.~\eqref{eq_NLDH_neutrality} expresses the neutrality of the system.


In itself, the NLDH model of ion-free-electron plasma is of no use, since it leads to the Coulomb collapse. However, in the context of non-ideality corrections, the accounting for interactions in the ion-free-electron plasma is often performed using two more-approximate models based on classical mechanics: the Debye-H\"{u}ckel model and the ion-sphere model. Each of these models is based on a different hypothesis, both of them allowing to circumvent the Coulomb catastrophe.

\subsection{Debye-H\"{u}ckel model}

The Debye-H\"{u}ckel model \cite{DebyeHuckel23} (DH) is based on a linearization with respect to the mean-field potential in Eq.~\eqref{eq_NLDH_Boltzmann}.
\begin{align}
n_{\alpha,\gamma}(r)=n_\gamma \left( 1-\beta Z_\gamma v_{\alpha}(r) \right)
\label{eq_DH_linearization}
\end{align}

This linerization both allows to have an analytical solution and to circumvent the Coulomb catastrophe. It is valid in the weak-coupling limit, but is in fact strongly unphysical in the vicinity of the nucleus at any conditions. This yields an unphysical singularity at the origin in the Debye-H\"{u}ckel correlation functions, which can be seen as a consequence of circumventing the Coulomb catastrophe.

The analytical solution of Eqs~\eqref{eq_NLDH_Percus},\eqref{eq_NLDH_Poisson},\eqref{eq_NLDH_boundary},\eqref{eq_NLDH_neutrality}, and \eqref{eq_DH_linearization} is
\begin{align}
&v_{\alpha}^\text{DH}(r)=Z^*_\alpha \ee^2 \frac{e^{-r/\lambda_\text{D}}}{r}\\
&g_{\alpha,\gamma}^\text{DH}(r)=1-\beta Z^*_\alpha Z^*_\gamma \ee^2 \frac{e^{-r/\lambda_\text{D}}}{r}\label{eq_DH_solution}
\end{align}
where $\lambda_\text{D}=\left( 4\pi \beta \ee^2 \sum_{\alpha=1}^{M,\text{e}}n_\alpha Z_\alpha^2 \right)^{-1/2}$ is the Debye length.

The mean-field potential applied on the central particle of species $\alpha$, due to the other particles is the divergence-free mean-field potential at the origin:
\begin{align}
v_{\alpha}^{*\text{DH}}=v_{\alpha}^\text{DH}(r)-\frac{Z_\alpha \ee^2}{r}
=-\frac{Z_\alpha \ee^2}{\lambda_\text{D}}
\label{eq_DH_potential}
\end{align}
The excess free energy can be calculated either using an integration over the inverse temperature (see \cite{LandauStatisticalPhysics}), or using the Debye-Kirkwood  charging method \cite{Kirkwood35} (see, for instance, \cite{Piron19a} for an application to the DH model with arbitrary potential).
\begin{align}
F_{\text{ex}}^\text{DH}=-\frac{1}{12\pi\beta n_\text{i}\lambda_\text{D}^3}
\label{eq_DH_free_energy}
\end{align}
From this free energy, the excess contribution to the pressure and internal energy are readily obtained, and it can be shown that the model fulfills the virial theorem. Thus, adding the excess free-energy $F_{\text{ex}}^\text{DH}$ to the Saha model, as in Eq.~\eqref{eq_Saha_correction_Fex}, preserves the virial theorem Eq.~\eqref{eq_virial_theorem}.

The free energy can also be used to calculate the chemical potentials
\begin{align}
\mu_{\text{ex},\alpha}^\text{DH}=-\frac{1}{2}\frac{Z^{*\,2}_\alpha \ee^2}{\lambda_\text{D}}
\label{eq_DH_chem_pot}
\end{align}
which yields the following correction to the ionization potentials (see, for instance, \cite{Griem62}):
\begin{align}
&\Delta I_\alpha^\text{DH} = -\frac{(Z^*_\alpha + 1)\ee^2}{\lambda_\text{D}}
\end{align}
The latter correction is equivalent to applying the correction $v_{\alpha}^{*\text{DH}}$ using the upper charge state, as in \cite{StewartPyatt66}.

In the average-atom context, the excess free energy of Eq.~\eqref{eq_AAII_Fex_correction} is:
\begin{align}
\dot{F}_{\text{ex}}^\text{DH}=-\frac{1}{12\pi\beta n_\text{i}}\left(4\pi \ee^2 \beta \left(n_\text{e}+n_\text{i}\frac{n_\text{e}^2}{n_\text{i}^2}\right)\right)^{3/2}
\label{eq_DH_free_energy_AA}
\end{align}
Accouting for this interaction correction leads to shifting the 1-electron eigenvalues as follows:
\begin{align}
&\varepsilon_\xi^* =\varepsilon_\xi + \frac{(Z^*+1/2)\ee^2}{\lambda_\text{D}}
\label{eq_DH_AA_correction}
\end{align}


\subsection{Ion-sphere model from a classical-plasma perspective}

\begin{figure}[ht]
\centerline{\includegraphics[width=8cm]{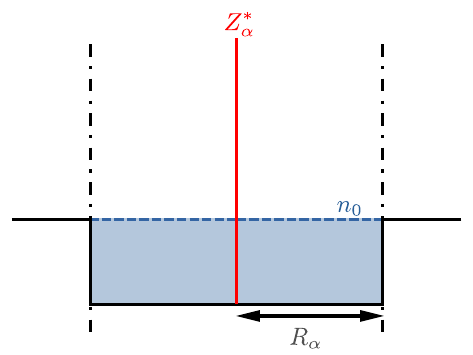}}
\caption{Schematic picture of the ion-sphere model. The electron density of the uniform background corresponds to the free-electron density of the plasma.
The radius of the sphere is such as to neutralize the central ion.
\label{fig_schema_ion_sphere}}
\end{figure}

In order to apply the notion of IPD to strongly coupled plasma, one often resorts to the ``ion-sphere'' (IS) model (see e.g. \cite{StewartPyatt66}). In this model, one considers a point-like ion placed at the center of a sphere filled only with a uniform density of free electrons, with which it interacts (see Fig.~\ref{fig_schema_ion_sphere}).
The uniform density of free electrons correspond to the mean free-electron density of the plasma $n_\text{e}=\sum_\alpha n_\alpha Z_\alpha^*$, while the radius of the sphere is such that the ion sphere is neutral: 
\begin{align}
R_\alpha=\left(\frac{3 Z_\alpha^*}{4\pi n_\text{e}}\right)^{1/3}
\end{align}
In the case of an average ion, $n_\text{e}=n_\text{i} Z^*$, the ion charge $Z_\alpha^*$ is replaced by $Z^*$ and the sphere radius is just the Wigner-Seitz radius $R_\text{WS}=(3 /(4\pi n_\text{i}))^{1/3}$.

From this model, one may evaluate the interaction energy of the central ion with the surrounding electrons in the sphere. This yields the correction $v_\alpha^{*\,\text{IS}}$ to the energy of the $\alpha$ species.
\begin{align}
v_\alpha^{*\,\text{IS}}
=-\frac{3}{2}\frac{Z^{*}_\alpha \ee^2}{R_\alpha}
\end{align}
One may also evaluate the total interaction energy of the ion sphere, which corresponds to the energy added if one adds an ion to the system, and thus yields the correction to the chemical potential.
\begin{align}
\mu_{\text{ex},\alpha}^\text{IS} =-\frac{9}{10}\frac{Z^{*\,2}_\alpha \ee^2}{R_\alpha}
\end{align}

On the interpretation of this model, two different physical pictures may be put forward. In the first, the plasma is seen as a highly-structured set of neutral spheres, resembling somehow to a solid-state situation. This interpretation is given, for instance, in \cite{StewartPyatt66}. Another interpretation is that the medium surrounding the ion may be split in two regions: a spherical statistical cavity in which other ions do not enter, and a uniform neutral plasma beyond the cavity (physical picture illustrated on Fig.~\ref{fig_schema_ion_sphere}). This resembles more to a liquid-state situation, and we will elaborate on the latter picture hereafter.

In the ion-sphere model, the founding hypothesis that allows to circumvent the Coulomb catastrophe is to neglect the polarization of the free electrons, which is driven by the attractive long range potential. Free electrons then constitutes a uniform background neutralizing the ions. If multiple ion species are considered, the system corresponds to the multi-component classical plasma (MCP) system. The Hamiltonian of such a system is
\begin{align}
&\mathcal{H}_\text{MCP}(\{\vec{p}_{\alpha,j},\vec{r}_{\alpha,j}\})=
\mathcal{H}_\text{id}(\{\vec{p}_{\alpha,j}\})
+W_\text{MCP}(\{\vec{r}_{\alpha,j}\})\\
&W_\text{MCP}(\{\vec{r}_{\alpha,j}\})
=\frac{\ee^2}{2}
\sum_{\alpha=1}^{M}\sum_{j=1}^{N_\alpha}
\mathop{ \sum_{\gamma=1}^{M}\sum_{k=1}^{N_\gamma} }_{(\gamma,k)\neq (\alpha,j)}
\frac{Z^*_\alpha Z^*_\gamma}{| \vec{r}_{\alpha,j} - \vec{r}_{\gamma,k} |}
-\ee^2\sum_{\alpha=1}^{M}\sum_{j=1}^{N_\alpha}\int_V d^3r\left\{\frac{n_\text{e} Z^*_\alpha}{| \vec{r}_{\alpha,j} - \vec{r} |}\right\}
+\frac{\ee^2}{2}\int_V d^3r d^3r'\left\{\frac{n_\text{e}^2}{| \vec{r}' - \vec{r} |}\right\}
\end{align}
where the sums only run over the $M$ ions species. If only one ion species is considered, the system corresponds to the one-component plasma (OCP).

Rigorously speaking, such rigid electron background hypothesis is applicable only in case of electron-electron and ion-electron coupling much weaker than the ion-ion coupling. Such conditions are found only when free electrons are strongly degenerate. Yet, this model is often used in practice to address the qualitative behavior of moderately- or strongly-coupled plasmas, as the only classical model available.

The OCP/MCP system may be addressed in the thermodynamic limit of the canonical ensemble either using methods of simulation \cite{Brush66,Hansen73} or statistical physics models (e.g. hypernetted chain \cite{Springer73,Ng74}, Percus-Yevick \cite{Carley63,Carley67}...). Among the statistical-physics models, the mean-field approach yields the nonlinear Debye-H\"{u}ckel model (or Poisson-Boltzmann model) of the OCP/MCP.

\begin{align}
&n_\gamma h_{\alpha,\gamma}(r)=n_\gamma(g_{\alpha,\gamma}(r)-1)=n_{\alpha,\gamma}(r)-n_\gamma
\label{eq_NLDHOCP_Percus}\\
&\nabla^2 v_{\alpha}(r)=-4\pi \ee^2\sum_{\gamma=1}^{M,\text{e}}Z_\gamma^* (n_{\alpha,\gamma}(r)-n_\gamma) \label{eq_NLDHOCP_Poisson}\\
&\lim_{r\rightarrow 0}v_{\alpha}(r)=\frac{Z_\alpha^* \ee^2}{r} \label{eq_NLDHOCP_boundary}\\
&\lim_{r\rightarrow \infty}v_{\alpha}(r)=0 
\label{eq_NLDHOCP_boundary_inf}\\
&n_{\alpha,\gamma}(r)=n_\gamma e^{-\beta Z_\gamma^* v_{\alpha}(r)}
\label{eq_NLDHOCP_Boltzmann}\\
&\sum_{\gamma=1}^{M}n_\gamma Z_\gamma^*=n_\text{e}
\label{eq_NLDHOCP_neutrality}
\end{align}

In the weak coupling limit $\Gamma_{\alpha,\gamma}\rightarrow 0$, one can perform the Debye-H\"{u}ckel linearization, replacing Eq.~\eqref{eq_NLDHOCP_Boltzmann} by Eq.~\eqref{eq_DH_linearization}. This leads to the DH model of the OCP/MCP, that is, a DH model in which polarization of free electrons is neglected.

In the strong-coupling limit $\Gamma_{\alpha,\gamma}\rightarrow\infty$ of this model, the pair distribution functions $g_{\alpha,\gamma}(r)$ take the form of Heaviside functions, yielding spherical, statistical cavities around each ion:
\begin{align}
g_{\alpha,\gamma}(r)\rightarrow \theta(r-R_{\alpha,\gamma})
\end{align} 

\begin{figure}[ht]
\centerline{\includegraphics[width=8cm]{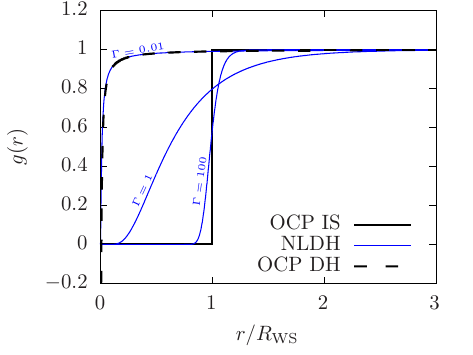}}
\caption{Pair distribution function $g(r)$ from the NLDH model of the OCP for various values of $\Gamma$. Comparison with $g(r)$ obtained from the Debye-H\"{u}ckel model of the OCP at $\Gamma=0.01$, and with the ion-sphere assumption for $g(r)$.
\label{fig_OCP_NLDH_IS_g}}
\end{figure}

In the OCP case, there is only one ion-ion pair distribution function to consider and the mean density of ion charge $\rho(r)$ around an ion directly becomes that of the ion-sphere with radius $R_\text{WS}$.
\begin{align}
\rho(r)
=n_\text{i} Z^* \ee \theta(r-R_\text{WS})
=n_\text{e} \ee \theta(r-R_\text{WS})
\end{align}

\begin{figure}[ht]
\centerline{\includegraphics[width=17cm]{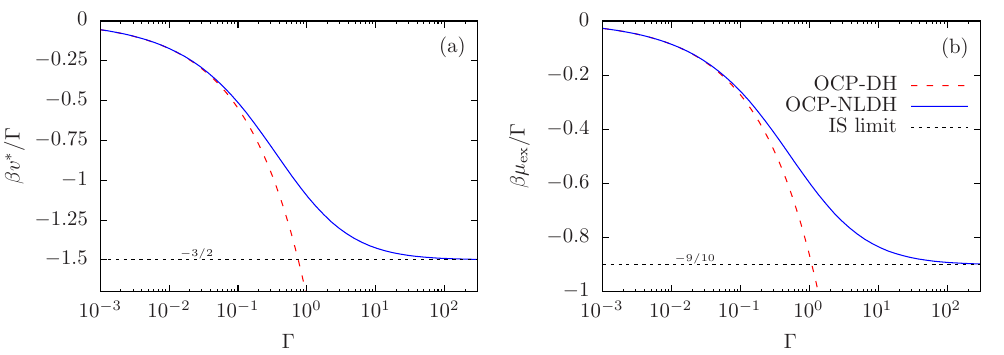}}
\caption{Chemical potential and divergence-free average potential at the origin for the Nonlinear Debye-H\"{u}ckel model of the one-component plasma.
\label{fig_OCP_NLDH_IS_mu}}
\end{figure}

Figure~\ref{fig_OCP_NLDH_IS_g} shows how the OCP pair distribution function obtained from the NLDH model changes from the DH shape to the IS shape, when $\Gamma$ is increased. Consistently, figure~\ref{fig_OCP_NLDH_IS_mu} displays the values of the divergence-free mean-field potential at the origin $v^*$ and of the excess chemical potential $\mu_\text{ex}$ as functions of $\Gamma$. One clearly sees how these two quantities goes from the DH behavior (in the sense of the OCP) to the IS limiting values.

\begin{figure}[ht]
\centerline{\includegraphics[width=8cm]{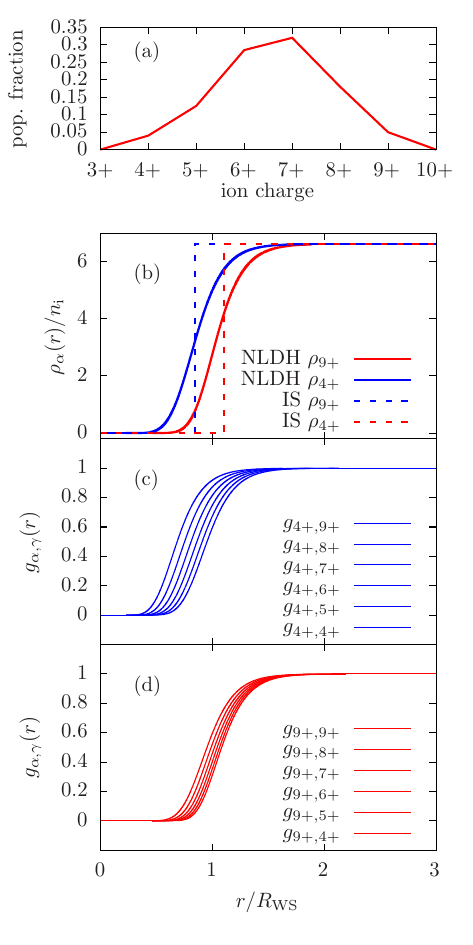}}
\caption{Example of a 6-component classical plasma, typical of the charge-state distribution of an iron plasma at solid density (7.8\,g.cm$^{-3}$) and 40-eV temperature, after \cite{Piron13} (a). Ion-ion coupling of significantly-populated ions spans from $\Gamma_{4+,4+}=1.4$ to $\Gamma_{9+,9+}=7.6$.
Plots of the average ion charge density $\rho_\alpha(r)$ around an ion of species $\alpha$ (b) and of the related pair distribution functions $g_{\alpha,\gamma}(r)$ (c, d) obtained using the NLDH model of the MCP, for two charge states. Comparison with the IS approximation to $\rho_\alpha(r)$ is shown in (b).
\label{fig_OCP_NLDH_IS_g_MCP}}
\end{figure}

In the MCP case, the mean density of ion charge $\rho_\alpha(r)$ around a given ion of species $\alpha$
results from the sum of the multiple pair distribution functions $g_{\alpha,\gamma}$. In the $\Gamma_{\alpha,\gamma}\rightarrow\infty$ limit, it ultimately tends to the ion sphere form:
\begin{align}
\rho_{\alpha}(r)=\sum_{\gamma}n_\gamma Z^*_\gamma \ee \theta(r-R_{\alpha})
\approx n_\text{e} \ee(1-\theta(r-R_\alpha))
\end{align} 

As an example, we present in Fig.~\ref{fig_OCP_NLDH_IS_g_MCP} the case of a 6-component classical plasma corresponding to the charge-state distribution obtained from the DCA model \cite{Piron13} for an iron plasma at solid density and 40-eV temperature. Ion coupling in this case is moderate to strong, going from $\Gamma_{4+,4+}=1.4$ to $\Gamma_{9+,9+}=7.6$. 

Let us mention that the hypothesis of an heaviside-shaped cavity with a radius varying so as to ensure neutrality for each charge state is precisely the hypothesis made in the model described in \cite{Piron13}, and the present discussion gives the physical picture underlying this hypothesis. 

\begin{figure}[ht]
\centerline{\includegraphics[width=17cm]{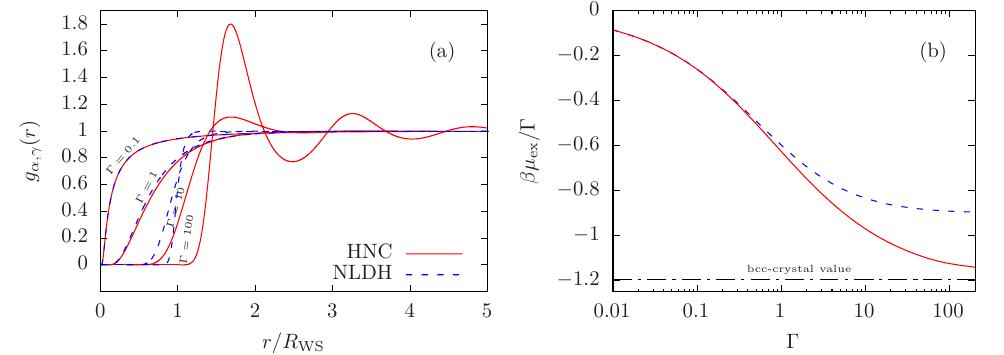}}
\caption{Comparison between the NLDH and HNC models of the OCP. Pair distribution functions for various values of $\Gamma$ (a) and excess chemical potential as a function of $\Gamma$ (b).
\label{fig_OCP_NLDH_HNC}}
\end{figure}

The problem of bridging the gap between the DH model of ion-free-electron plasma and the IS model of the OCP is notably adressed in \cite{StewartPyatt66}, using an approach inspired from the Thomas-Fermi model (see next section). However, in this paper, the model is used to describe the electron cloud around an ion, rather than the electronic structure of the ion itself. In this approach, the Thomas-Fermi electron density enables a gradual switching to the rigid-background behavior of the electrons, when the degree of electron degeneracy is increased.


Finally, whereas the DH model of ion-free-electron plasma may be seen as rigorously valid in the limit of weak coupling, the ion-sphere model may only be related to the mean-field approach of the OCP/MCP model. In case of strong coupling, the effect of correlations, beyond the reach of the mean-field approximation, becomes important. Models that accounts for these effects, such as the hypernetted-chain model (HNC) \cite{Morita58,Morita59}, exhibit a different behavior of the pair distribution functions at high $\Gamma$, as well as a different limit for the chemical potential. Figure~\ref{fig_OCP_NLDH_HNC} illustrates these differences between the NLDH and HNC model of the OCP, as regards both the pair distribution function and the excess chemical potential.

One may then put forward the other interpretation of the ion-sphere model: a highly-ordered, periodical set of neutral spheres, which may consistute a good approximation to the lattice situation. Indeed, it is known that the OCP system crystallizes at $\Gamma$ close to 175, yielding a Wigner crystal having a body-centered-cubic lattice (bcc, see \cite{VanHorn69,Hansen72,Slattery82}). For the bcc crystal of an OCP, at zero temperature, we have \cite{Slattery82}:
\begin{align}
&\beta U_0=-0.895929\,\Gamma\\
&\beta\mu=\frac{4}{3} \beta U_0
\end{align}
where $U_0$ is the internal energy per ion.

As may be seen on Fig.~\ref{fig_OCP_NLDH_HNC}, a statistical model accounting for correlations beyond the mean-field approximation, such as HNC, yields results much closer to those of the OCP bcc lattice than does the IS model. In fact, the close proximity in a large range of $\Gamma$ between thermodynamic quantities stemming from the fluid phase and those from the bcc crystal is among the challenges in the precise determination of the OCP phase transition.




\subsection{Suppression of bound states, screening and limitations of the nonideality corrections}
For the most weakly-bound states (both in the sense of many-electron states or in the sense of 1-electron orbitals), the non-ideality correction to the energy may be of the same order as the energy itself, or even greater. The question then is: how to account for these states ? This question is in fact deeply connected to the issue of truncating the set of levels in the Saha partition function, briefly mentioned in section~\ref{sec_ideal_plasmas}.

In \cite{Herzfeld16}, Herzfeld employs the heuristic argument of the mean radius of an hydrogenic orbital being of the same order of half the interparticle distance, in order to truncate the set of bound states accounted for in the Saha equilibrium for hydrogen. Qualitatively, this corresponds to limiting the spatial extension of existing states to the size of an ion sphere. In \cite{Fermi24}, Fermi elaborate on this idea, introducing in the free energy an excluded-volume term associated with the volume occupied by the various hydrogenic states. Minimizing the free energy, he obtains that the population of a state drops when the total volume does not withstand its volume times the population of the most populated state.

In \cite{Larkin60}, Larkin relates the truncation to an assumed Debye-H\"{u}ckel decay of the 1-electron effective potential. Qualitatively, this corresponds to limiting the spatial extension of existing states to the Debye length. However, without a unified treatment for both the electronic structure and the screening in the whole plasma, the argument for using the Debye-H\"{u}ckel potential in the ion electronic structure is heuristical. It is also possible to recover the Larkin truncation by introducing interactions through the virial expansion, which leads to the convergent Planck-Larkin partition function \cite{Rogers86}. However this rigorous perturbative analysis is in practice restricted to the Debye-H\"{uckel} limit. 

In \cite{StewartPyatt66}, Stewart and Pyatt recommend to suppress any level for which the ionization-potential correction is greater than the level energy.

From a mathematical point of view, one can relate the suppression of bound states to the range of the 1-electron effective potential. Calculating the atomic orbitals in a finite-range potential instead of a potential having a Coulomb tail implies qualitative changes on the ion electronic structure. 

Let us define the phase-shift $\Delta_{k,\ell}(r)$ of a continuum radial wavefunction $P_{k,\ell}(r)$ with respect to the regular Bessel function:
\begin{align}
&\varphi_{k,\ell,m}(\vec{r})=\frac{P_{k,\ell}(r)}{r}Y_{\ell,m_\ell}(\hat{r})
\\
&P_{k,\ell}(r) = A_{k,\ell}(r)k.r\left( \cos(\Delta_{k,\ell}(r)) J_\ell(k.r+\Delta_\ell)
-\sin(\Delta_{k,\ell}(r)) Y_\ell(k.r+\Delta_\ell) \right)
\end{align}
where $J_\ell$ and $Y_\ell$ are the spherical Bessel functions, regular and singular at 0, respectively.

For a finite-range potential, the Schr\"{o}dinger equation tends to the Bessel equation far from the origin, and the phase shift has a finite asymptotic value.
\begin{align}
\lim_{r\rightarrow\infty}\Delta_{k,\ell}(r)=\Delta_{k,\ell}
\end{align}
On the contrary, for a Coulomb-tail potential, the solutions of the Schr\"{o}dinger equation tends to the Coulomb wavefunctions, yielding:
\begin{align}
\lim_{r\rightarrow\infty}\Delta_{k,\ell}(r)\propto \log(kr)
\end{align}

A theorem due to Levinson \cite{Levinson49} (see \cite{Zhong-Qi06} for a presentation better suited to the present application) relates the asymptotic phase-shift at zero energy to the number of existing bound states $N_\text{b}=\lim_{k\rightarrow 0}\Delta_{k,\ell}/\pi$. When the phase shift diverges, the number of bound states is infinite, whereas in case of a finite asymptotic value, the number of bound states is finite. Thus, for a potential having finite range, the number of discrete orbitals is finite \cite{Bargmann52}. The same reasoning is at the origin of the result of \cite{BethUhlenbeck37} for the second virial coefficient, leading to the convergent Planck-Larkin partition function.

\begin{figure}[ht]
\centerline{\includegraphics[width=10cm]{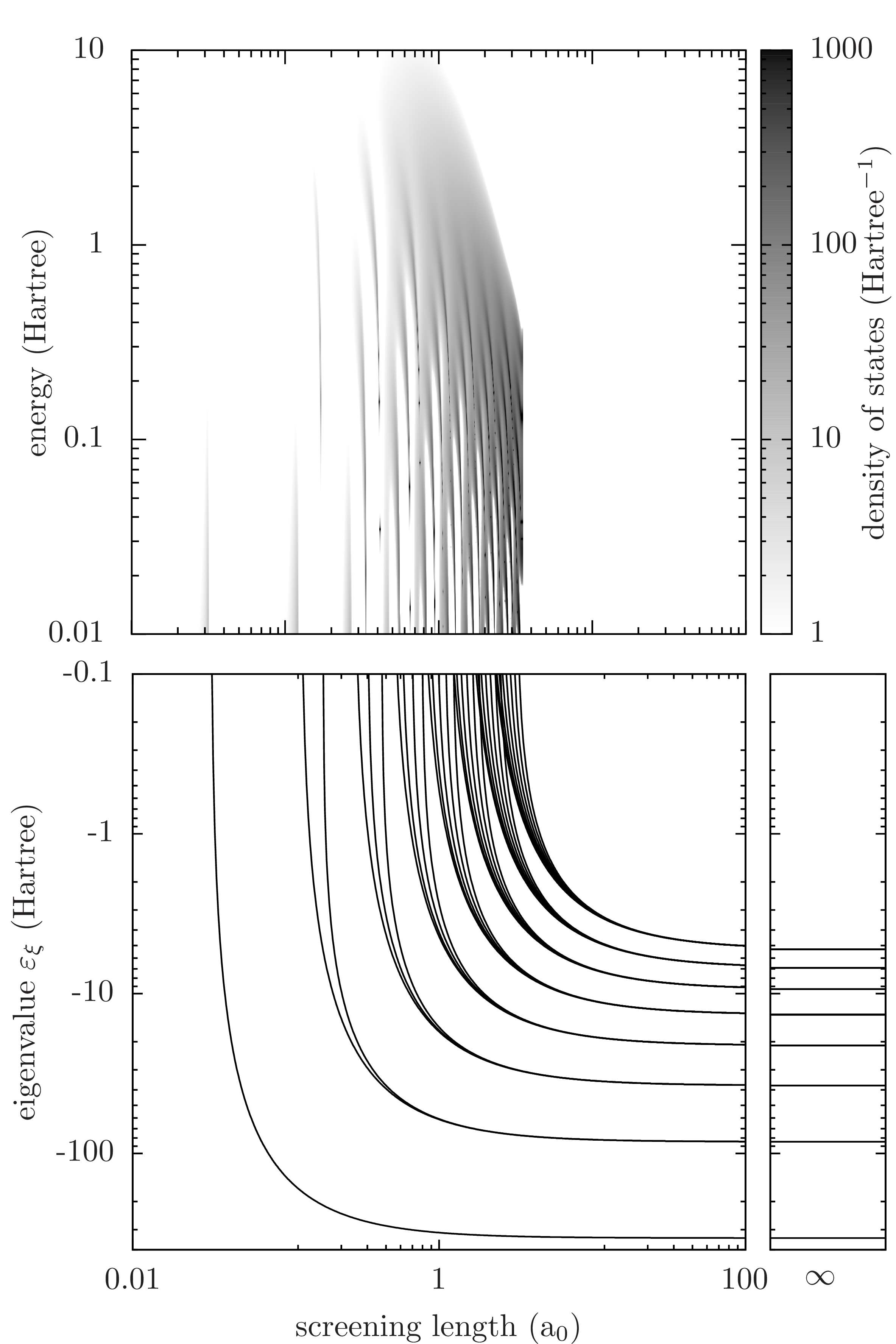}}
\caption{Electron in a screened Coulomb potential with charge $Z=26$. Eigenvalues as functions of the screening length, for principal quantum numbers up to $n=8$, and ion contribution to the density of states $\Delta\rho(\varepsilon)$ of Eq.~\eqref{eq_phaseshift_dos}, showing the corresponding resonances. 
\label{fig_Yukawa_screening}}
\end{figure}

If the screened potential corresponds to the atomic potential of an ion, and if wavefunctions are normalized in the whole space, then one may relate the phase shift to an ion contribution $\Delta\rho(\varepsilon)$ to the density of states as follows:
\begin{align}
\Delta\rho(\varepsilon)=2\sum_\ell (2\ell+1)\frac{1}{\pi}\frac{\partial\Delta_{k,\ell}}{\partial \varepsilon}
\label{eq_phaseshift_dos}
\end{align}
As an elementary illustration of the effect of a finite-range potential, figure~\ref{fig_Yukawa_screening} present the 1-electron eigenvalues and the $\Delta\rho(\varepsilon)$ contribution to the density of states for a screened Coulomb potential, with charge $Z=26$, as a function of the screening length. For infinite screening length, one recovers the usual Coulomb eigenvalues, with the accidental $\ell$-degeneracy. As the screening length is decreased, one can see how the accidental degeneracy is removed and how the eigenvalues are gradually shifted towards the continuum until the removal of the orbital from the discrete spectrum. Once the orbital is removed from the discrete spectrum, one can see how it it compensated by a resonance in the continuum, which gradually spreads out as screening length is further decreased.

However, for any elementary change of the potential that yields the removal of a discrete bound state, any observable has to remain continuous. Otherwise, it would be basis-dependent. This necessarily implies the compensation of the contribution of the removed discrete state by an equivalent contribution of the continuous spectrum, related to a resonance in the density of states, which is a property of the basis (see, for instance \cite{More85}).

One may raise the point that, if one sees the non-ideality correction as a lowering of the continuum, then the intersection of this lowered continuum boundary with the isolated ion potential sets a restriction on the range of the potential, yielding the truncation of the bound state spectrum. This is another qualitative standpoint, fully consistent with that of suppressing the states whose energy lies above the lowered continuum. However, in this case, the contribution of resonances in the continuum is disregarded. Even assuming that the continuum-lowering model is yielding the correct energy shift, a sharp suppression of the bound state does not correspond to what stems from a screened potential.

In practice, the continuum-lowering argument or the introduction of Planck-Larkin partition functions suffice to justify the suppression of weakly-bound states, which have negligible populations, and do not yield significant contribution of the corresponding resonances. This suppression enables convergence of the Saha partition function. However, when it comes to suppressing populated many-electron states or orbitals, this method is no more valid, and in fact the whole point-like-ion hypothesis used in the treatment of interactions breaks down.

A proper answer to the problem requires an accounting for the interactions among particles of the plasma directly in the calculation of the ion electronic structure. As far as possible, such a calculation should account for both the polarization of free electrons around the ions, and the interactions of ions with their neighbours. This is the purpose of models of pressure-ionized plasmas.

\begin{figure}[ht]
\centerline{
\includegraphics[width=8cm]{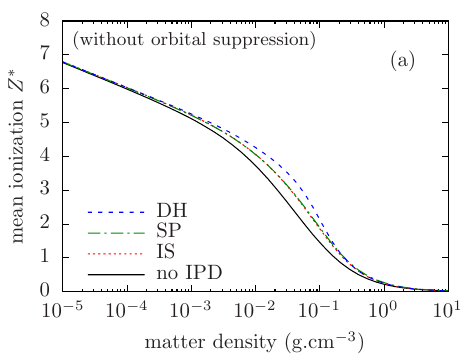}
\hspace{1cm}
\includegraphics[width=8cm]{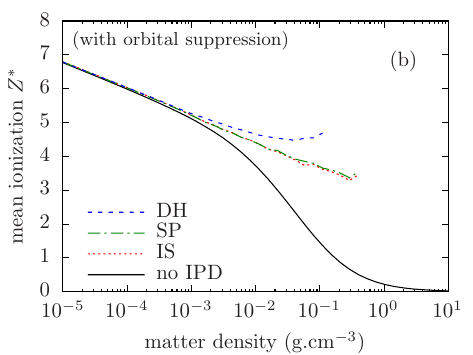}
}
\caption{Mean ionization of a silicon plasma at 20 eV temperature, as a function of matter density. Comparison between average-atom model of isolated ion without any continuum lowering (no IPD, principal quantum number limited to $n=8$), with Debye-H\"{u}ckel continuum lowering (DH), with ion-sphere continuum lowering (IS), and with Stewart-Pyatt continuum lowering (SP). Comparisons are shown both without suppression of bound orbitals (a) and with suppression (b). In the latter case, the curves stop where suppression of a subshell having more than 10\% of the electrons occurs (regime of significant pressure ionization).
\label{fig_DH_IS_suppr}}
\end{figure}

Figure~\ref{fig_DH_IS_suppr} shows the effect of various ionization potential depression models, either disregarding or performing the suppression of bound orbitals. Looking at Fig.~\ref{fig_DH_IS_suppr}a, one can see that without suppressing any orbital, the effect on the mean ionzation remains moderate, Debye-H\"{u}ckel model yielding the largest effect. In the case considered, Stewart-Pyatt formula leads to results which are close to those of the ion-sphere model. Looking at Fig.~\ref{fig_DH_IS_suppr}b, one can see that performing the associated suppression of bound orbitals, the impact on the mean ionzation is much more pronounced. In fact, most of the effect of accounting for non-ideality corrections is in the modification of the partition function.

\section{Atomic models of pressure-ionized plasmas\label{sec_pressure_ionized_plasmas}}
Whenever the interactions of the ions with the surrounding ions and free electrons has an impact on the electronic structure, we will speak of ``pressure-ionized'' plasma. In such a case, the picture of a plasma of point-like ions and free electrons breaks down. It is then required to properly account for the surrounding particles in the modeling of the internal electronic structure of each ion. In the following, we will focus on the average-atom description of the plasma, since it offers a simpler framework for such modeling. However, most of the presented models can be extended to a detailed description of the plasma.

The modeling of dense, pressure-ionized plasmas has historically been addressed through self-consistent-field models of the ion electronic structure, including all electrons (bound and continuum) and accounting for the surrounding ions through the notion of a Wigner-Seitz cavity. These models focus on the electronic structure around a bare nucleus, and depart from the formalism of correlations in the plasma. 
Being models of the ion electronic structure, all these models necessarily rely, to some extent, on quantum mechanics for the electrons, and thus avoid the Coulomb catastrophe.

Depending on the model, the WS cavity is seen either as a neutral spherical cell in which the ion is enclosed (ion-in-cell picture), or as a statistical cavity within which surrounding ions do not enter, and beyond which they are uniformly distributed (ion-in-jellium picture). 

A common feature of these models of pressure-ionized plasmas is that the resulting atomic potential has finite range, and thus naturally leads to a finite number of bound states.

\subsection{Thomas-Fermi ion-in-cell model}

\begin{figure}[ht]
\centerline{\includegraphics[width=8cm]{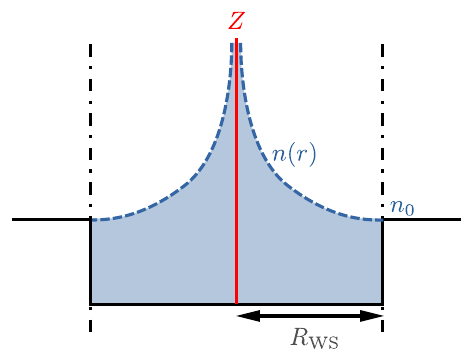}}
\caption{Schematic picture of the Thomas-Fermi model. 
\label{fig_schema_TF}}
\end{figure}

The Thomas-Fermi (TF) model is a semiclassical mean-field model of the ion electronic structure, accounting for all electrons and for the surrounding ions through the notion of an ion cell. Application of the TF model at finite temperature to dense plasmas was first proposed in \cite{Feynman49}. Numerical solution of the TF set of equations was given in \cite{Latter55}. The equations of the TF model are:
\begin{align}
&\nabla^2 v_\text{el}(r)=-4\pi \ee^2 n(r) \label{eq_TF_SCF}\\
&\lim_{r\rightarrow 0}v_\text{el}(r)=-\frac{Z\ee^2}{r} \label{eq_TF_boundary_0}\\
&v_\text{el}(R_\text{WS})=0 \label{eq_TF_boundary_WS}\\
&\int_\text{WS} d^3r\left\{n(r)\right\}=Z \label{eq_TF_neutrality_WS}\\
&n(r)=\frac{4}{\sqrt{\pi}\Lambda_\text{e}^3}I_{1/2}\left(\beta\left(\mu^\text{F}_\text{e}(n_\text{e})-v_\text{el}(r)\right)\right) \label{eq_TF_density}
\end{align}
where the WS denotes that the integral is performed only within the WS sphere.

Eq.~\eqref{eq_TF_SCF} is the Poisson equation. Eq.~\eqref{eq_TF_boundary_0}, \eqref{eq_TF_boundary_WS} are the boundary conditions at the origin and at the WS radius, respectively. The latter sets the reference of the energies.
Eq.~\eqref{eq_TF_neutrality_WS} is the condition of neutrality of the ion sphere. Eq.~\eqref{eq_TF_density} corresponds to the local-ideal-Fermi-gas hypothesis.
Eq.~\eqref{eq_TF_SCF}, together with Eq.~\eqref{eq_TF_density} imply a mean-field approximation.

An oft-used extension of the TF model consists in adding a local exchange, or an exchange-correlation contribution to the electrostatic potential. This is called the Thomas-Fermi-Dirac model, refering to \cite{Dirac30} in which a local exchange term was derived. In this case, Eq.~\eqref{eq_TF_density} in replaced by:
\begin{align}
n(r)
&=\frac{4}{\sqrt{\pi}\Lambda_\text{e}^3}I_{1/2}\left(\beta\left(\mu^\text{F}_\text{e}(n_\text{e})-v_\text{el}(r)-\mu_\text{xc}(n(r))\right)\right)
\label{eq_TFD_density}
\end{align}
with $\mu_\text{xc}={\partial f_\text{xc}(n)}/{\partial n}$ being the chemical potential associated to $f_\text{xc}$, an approximate exchange-correlation contribution to the free-energy per unit volume of a uniform electron gas.

Besides its heuristical setup, one may also derive the Thomas-Fermi(-Dirac) model from a variational principle. One approximate the free energy per ion as the free energy of a ion cell, filled with a electron gas, locally considered as an ideal Fermi gas of density $n(r)$.
\begin{align}
\dot{F}&\{\underline{n};n_\text{i},T\}
=\int_\text{WS} d^3r\left\{ f_\text{e}^\text{F}(n(r),T)+f_\text{xc}(n(r),T) \right\}
+\int_\text{WS} d^3r\left\{ -\frac{Z n(r) \ee^2}{r} +\frac{\ee^2}{2}\int_\text{WS} d^3r' \left\{ \frac{n(r)n(r)}{|\vec{r}-\vec{r}'|}\right\} \right\}
\end{align}
It is worth noting that this free energy does not include terms related to the ion motion or interactions. This is among the shortfalls of such kind of model, which only focus on the electronic structure of a central ion. In first approximation, an ion ideal-gas free energy contribution can be trivially added. However, a proper accounting for the ion-ion interactions in such a model is a far more difficult subject, on which we will elaborate later.
 
One performs the minimization of the free energy per ion while requiring the neutrality of the ion cell
\begin{align}
\dot{F}_\text{eq}(n_\text{i},T)
=\underset{\underline{n}}{\text{Min}}\,&\dot{F}\{\underline{n};n_\text{i},T\}
~\text{ s. t. } Z=\int_\text{WS} d^3r\left\{ n(r) \right\}
\label{eq_minimization_TF}
\end{align}
The result of this constrained minimization is equivalent to Eqs~\eqref{eq_TF_SCF},\eqref{eq_TF_boundary_0},\eqref{eq_TF_boundary_WS},\eqref{eq_TF_neutrality_WS},\eqref{eq_TFD_density}.

Starting from the equilibrium free energy, one can rigorously write the thermodynamic
quantities using the relevant derivatives. It can also be shown that the model fulfills the virial theorem \cite{Feynman49}.

In the TF model, the electron density is a local function of the potential. In fact, the hypothesis of having locally an ideal Fermi gas may be recovered from a local-density approximation to the quantum kinetic free energy of independent particles \cite{Kirzhnits57}. Since the electrostatic potential is zero at the WS radius, the density at the WS radius is equal to that obtained from the chemical potential:
\begin{align}
Z^*=\frac{n_\text{e}(R_\text{WS})}{n_\text{i}}=\frac{n_\text{e}}{n_\text{i}}
\label{eq_TF_mean_ioniz}
\end{align}

\begin{figure}[ht]
\centerline{
\includegraphics[width=8cm]{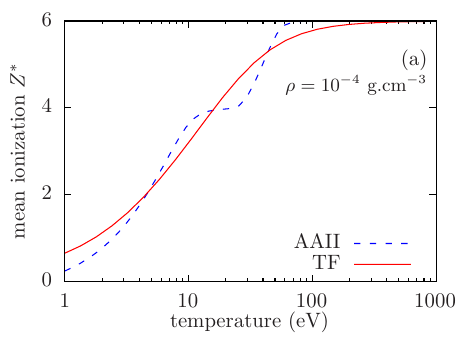}
\hspace{1cm}
\includegraphics[width=8cm]{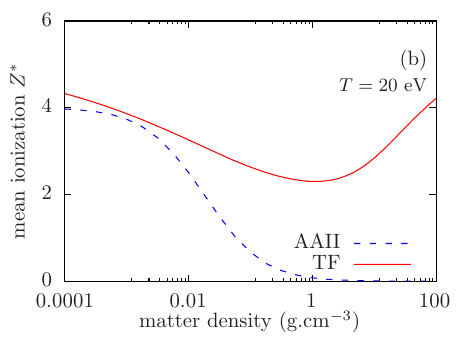}
}
\caption{Mean ionization of a C plasma along the $10^{-4}$-g.cm$^{-3}$-isochore (a), and along the 20-eV-isotherm (b). Comparison between the Thomas-Fermi model (TF) and the average-atom model of isolated ion (AAII).
\label{fig_TF_AAII}}
\end{figure}

Due to its semiclassical character, the TF model do not yield a shell structure in the sense of quantum mechanics. Consequently, there are no ionization plateau, as in the mean ionization of the quantum isolated ion. As an illustration, Fig.~\ref{fig_TF_AAII}a displays the TF mean ionization as a function of temperature for carbon at $10^{-4}$ g.cm$^{-3}$, compared to that of a quantum isolated ion.

On the other hand, in the TF model, pressure ionization is obtained through a squeezing of the ion cell when density is increased. As an example, Fig.~\ref{fig_TF_AAII}b shows the TF mean ionization as a function of matter density for carbon at a temperature of 20 eV.

In addition to being used in equation-of-state calculations, the TF model was also used for the calculation of radiative properties. Such calculations were performed either resorting to a heuristic use of the TF potential in orbital calculations \cite{Latter55b}, or from rigorous approaches to the dynamic semiclassical model \cite{Ball73,Ishikawa98b,Caizergues14}. In the latter case, the unphysical behavior of the TF electron density in the vicinity of the nucleus has strong consequences on the photoabsorption cross-section at high frequencies \cite{Ishikawa98b}. Moreover, the lack of shell structure implies the absence of line emission and absorption in the spectra. For the reasons given above, the need for a quantum extension of the TF model was quickly recognized.

Finally, let us remark that in the Thomas Fermi model, the equations are restricted to the WS cell. This renders the model versatile in the sense that it is rather insensitive to the modeling of the medium outside the WS sphere. On one hand, one may interpret the ion sphere as an element of a highly-ordered pile of neutral spheres. In \cite{Slater35}, the TF ion cell is used in this way, as an approximation to the polyhedral WS cell of a metal lattice.  On the other hand, one may interpret the ion cell as a statistical cavity surrounded by a homogeneous neutral plasma (jellium), as suggested later by Liberman in the context of is INFERNO model \cite{Liberman79}.
The coexistence of these two possible interpretations relates to the similar ambiguity of the physical picture underlying the ion-sphere model. This duality of interpretation left an imprint on the models proposed later as quantum extension to the TF model.

\subsection{Quantum ion-in-cell models}

\begin{figure}[ht]
\centerline{\includegraphics[width=8cm]{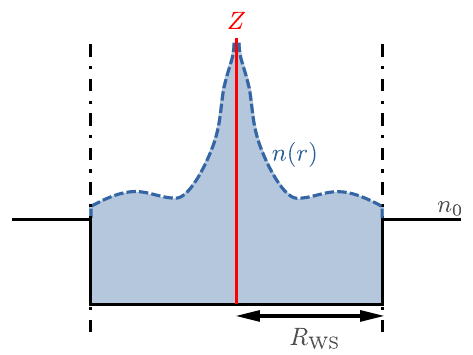}}
\caption{Schematic picture of Liberman's INFERNO model. The electron density is represented with a discontinuity at the WS radius, consistently with the interpretation proposed by Liberman of a homogeneous jellium surrounding the WS sphere.
\label{fig_schema_INFERNO}}
\end{figure}

Among the first quantum extension to the TF model was the model of Rozsnyai \cite{Rozsnyai72}. This model is based on the solid-state picture of the ion cell. The bound electrons are described resorting to energy bands, whose boundaries are obtained from the Wigner-Seitz cellular method (zeros of the wavefunction and of its derivative, see \cite{Wigner33,Wigner34} or the monograph \cite{AshcroftMermin}). Positive-energy spectrum (with respect to the effective potential at infinity) is approximated using the TF approach, with a restriction on the energy integration in order to only cover the classically-allowed range. The treatment of continuum electrons is therefore not consistent with that of bound electrons. In particular, the contributions of resonances or energy bands in the continuum are disregarded. However, because of the treatment of bound electrons through energy bands, the pressure ionization of a bound state occurs gradually and does not result in a proper discontinuity of observables.

A common variant of this model resorts to wavefunctions calculated with boundary conditions applied at infinity. In this case, the boundary condition is the exponential decay of the wavefunction, or, in practice, the matching onto localized zero-field solutions at the WS radius (third kind modified spherical Bessel function). This kind of model is, for instance used in \cite{Blenski00}. In this model, due to the semiclassical treatment of the continuum and the discrete nature of bound states, pressure ionization of a bound state results in a discontinuity of observables.

In practice, the equations of the latter model are the same as in the TF model, except that the electron density is partially calculated from quantum mechanics. Namely, one retains Eqs~\eqref{eq_TF_SCF},\eqref{eq_TF_boundary_0},\eqref{eq_TF_boundary_WS},\eqref{eq_TF_neutrality_WS}, whereas the electron density is given by:
\begin{align}
n(r)=&\sum_{\xi}p_\text{F}(\mu,T,\varepsilon_\xi)|\varphi_\xi(\vec{r})|^2
+\frac{4}{\sqrt{\pi}\Lambda_\text{e}^3}I_{1/2}^\text{inc.}\left(\beta\left(\mu^\text{F}_\text{e}(n_\text{e})-v_\text{el}(r)-\mu_\text{xc}(n(r))\right);
-\beta\left(v_\text{el}(r)+\mu_\text{xc}(n(r))\right)
\right)
\label{eq_Rozsnyai_density}
\end{align}
where the sum only runs over the discrete part of the spectrum, and where $I_{1/2}^\text{inc.}$ is the incomplete Fermi integral defined as follows:
\begin{align}
I_{1/2}^\text{inc.}\left(y;z\right)
=\int_z^\infty dx\left\{ \frac{x^{1/2}}{e^{x-y}+1} \right\}
\end{align}

Another slightly different variant of this model approximates the positive-energy spectrum using the non-degenerate limit of the 1-electron distribution instead of the Fermi-Dirac distribution \cite{Rosmej11}\footnote{In \cite{Rosmej11}, this model is called ``finite-temperature ion-sphere model'', whereas what we call in the present article ``ion-sphere model'' is called ``uniform electron gas model''.}.

The first fully-quantum model of the ion cell in a plasma was Liberman's model named ``INFERNO'' \cite{Liberman79,Liberman82,INFERNO}. Contrary to Rozsnyai, Liberman proposes the physical picture of an ion cell surrounded by a finite-temperature jellium, as sketched in Fig.~\ref{fig_schema_INFERNO}. A jellium is an homogeneous electron gas, neutralized by a homogeneous ion background. 

The equations of the INFERNO model are the same as in the TF model, with an electron density fully calculated from quantum mechanics. One thus keep Eqs~\eqref{eq_TF_SCF},\eqref{eq_TF_boundary_0},\eqref{eq_TF_boundary_WS},\eqref{eq_TF_neutrality_WS}, with an the electron density given by:
\begin{align}
&n(r)=\sum_{\xi}p_\text{F}(\mu,T,\varepsilon_\xi)|\varphi_\xi(\vec{r})|^2
\label{eq_inferno_density}
\end{align}
where the sum runs over both the discrete and the continuum part of the spectrum. For the continuum, the sum is to be understood as an integral over the momentum. Contrary to Rozsnyai model and its variants, the INFERNO model accounts for the resonances in the continuum. 

\begin{figure}[ht]
\centerline{\includegraphics[width=15cm]{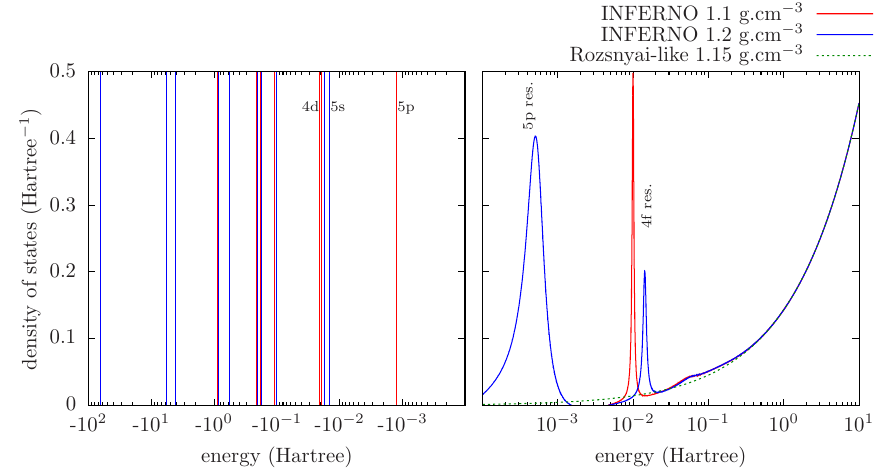}}
\caption{Density of states obtained from the INFERNO model in the case of silicon at 5 eV temperature and matter densities of 1.1 and 1.2 10$^{-2}$ g.cm$^{-3}$. In the negative energy ranges the Dirac distributions are represented by vertical lines. For the sake of comparison, the density of state stemming from the Rozsnyai-like model of Eq.~\eqref{eq_Rozsnyai_density}, at 5 eV temperature and 1.15 10$^{-2}$ g.cm$^{-3}$ is also shown, only in the positive energy range.
\label{fig_dos_INFERNO}
}
\end{figure}

Figure~\ref{fig_dos_INFERNO} shows the density of states obtained from the INFERNO model for silicon at 5 eV temperature and matter densities of 1.1 and 1.2 10$^{-2}$ g.cm$^{-3}$. At these conditions, one can observe resonances related to the delocalizations of the 5p and 4f subshells. In particular, between 1.1 and 1.2 10$^{-2}$ g.cm$^{-3}$, the 5p subshell gets pressure-ionized, yielding a sharp resonance in the continuous spectrum. In order to illustrate the lack of resonances in Rozsnyai-like models, we also display the density of states obtained when using Eq.~\eqref{eq_Rozsnyai_density} for the electron density, at similar plasma conditions. The discrete spectrum from the Rozsnyai-like model is not shown, to avoid obfuscation of the figure.

Consistently, with the picture of an ion cell surrounded by a neutral jellium, the boundary condition applied to the wavefunctions $\varphi_\xi$ at the WS radius is just the matching onto the zero-potential solution (linear combination of Bessel functions, defining a phase shift). Like in the TF model, the medium surrounding of the ion cell does not interact with the content of the ion cell. Consequently, the model can be formulated using equations restricted to the WS cell, the jellium surrounding the WS sphere playing no direct role in the model. 
 
The quantum electron density of Eq.~\eqref{eq_inferno_density} is a nonlocal functional of the self-consistent potential. As a consequence, even if the potential is zero at the WS radius, the electron density $n(R_\text{WS})$ in general differs from the electron density obtained from the chemical potential $n_\text{e}$. The latter corresponds to the asymptotic value $\lim_{r\rightarrow\infty}n(r)$.

This yields an ambiguity of definition of the mean ionization $Z^*$ which is closely related to the ambiguity in the physical interpretation of the model. Either the electron density has a discontinuity at the WS radius, or it is continuous but electrons of the jellium have a chemical potential different from those of the ion cell.

Connected to this interpretation issue is the problem of defining the pressure in the model (electron pressure at the WS boundary versus electron pressure stemming from a jellium of density $n_\text{e}$). More generally, due to the lack of variational formulation for this model, any thermodynamic quantity is defined heuristically, and may have more than one possible definition. Indeed, Liberman proposed two versions of his model (denoted A and T), differing in the region of integration for the free and internal energies \cite{Liberman79,Liberman82,INFERNO}. Thermodynamic consistency among these quantities is in general not assured. 

The sharp cut-off of the equations at the WS radius also implies that the virial theorem is not fulfilled. When trying to derive the virial theorem for the system, surface terms appears at the WS radius, which results in the impossibility for the model to fulfill the theorem (see, for instance \cite{PironPhD}).

Nevertheless, Rozsnyai's and Liberman's models are among the most often used when dealing with pressure-ionized plasma, both in their respective average-atom versions (see for instance \cite{Wilson06,Penicaud09}) or in modified version adapted to fixed configurations (see, for instance \cite{Blenski00}). To some degree of approximation, these models account for both the quantum shell structure of the ion and the pressure ionization phenomenon. Both also have relatively light computation costs, favored by the restriction of the equations to the WS cell. Of course, INFERNO involves a much higher computational cost than Rozsnyai's model, due to the quantum treatment of the continuum.

Moreover, a variant of Rozsnyai's model was used in \cite{Rosmej11} as the starting point to get an approximate, closed formula fitting the atomic potential. Such a fit for the atomic potential can be used to infer corrections to the isolated-ion energies, from perturbation theory \cite{Li12,Iglesias19}. Such an approach yields analytical formulae for the line shifts, which showed agreement with experimental measurments of line shifts in He-like ions at electron densities of the order of $10^{23}$ cm$^{-3}$ \cite{Li20}.

\subsection{Ion-in-jellium models}
\begin{figure}[ht]
\centerline{
\includegraphics[width=8cm]{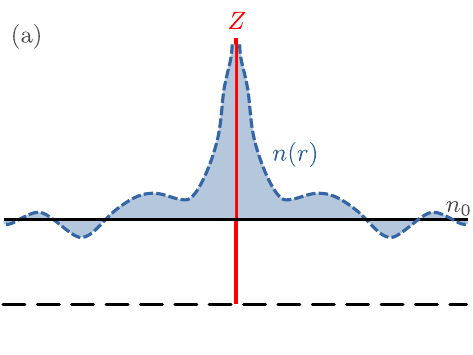}
\hspace{1cm}
\includegraphics[width=8cm]{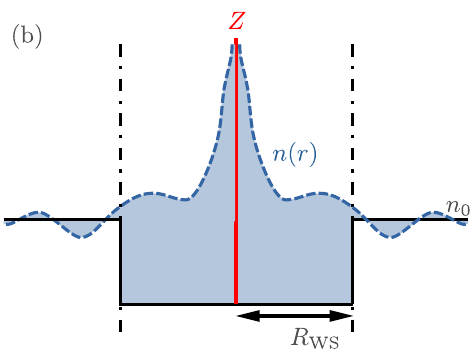}}
\caption{Schematic pictures of an impurity in a jellium (a), and of an ion-in-jellium model such as AJCI or VAAQP (b).
\label{fig_schema_ion_jellium}}
\end{figure}
Models of an impurity (or a defect) in a jellium were developped during the 70's in the context of solid-state physics \cite{Arponen73,Manninen75,Jena78}. In these models, the perturbation generated by the impurity may extend spatially far from its origin. There is no restriction to a particular cell (see Fig.~\ref{fig_schema_ion_jellium}a).

A first extension of the treatment of an impurity in a jellium to the modeling of an ion in a plasma was suggested by Perrot in the 90's, in his ``Atome dans le Jellium de Charge Impos\'ee'' model (AJCI, atom in a jellium with fixed charge) \cite{PerrotReport}. In his model, Perrot introduces a WS statistical cavity in the jellium, much like the picture proposed by Liberman. However, he also considers an ion extending in the whole space, rather than enclosed within a cell (see Fig.~\ref{fig_schema_ion_jellium}b). Consistently, the neutrality is assumed to hold in the whole space, rather than in the ion cell. In this model, the uniform ion background of the jellium surrounding the cavity interacts with the electron density, which asymptotically tends to the jellium density. This leads to the charge density:
\begin{align}
\ee \left( n(r)-n_\text{i}Z^*\theta(r-r_\text{WS}) \right)
=\ee \left( n(r)-n_\text{e}\theta(r-r_\text{WS}) \right)
\end{align} 

The AJCI model, like models of impurity in metals, resorts to a fixed jellium density $n_\text{e}$, given as an input to the model. It is also lacking a variational derivation. However, the notion of an ion extending beyond the WS sphere, up to infinity, allows in principle to solve the problem of surface terms in the virial theorem.

Starting from the founding ideas of the AJCI model, a model of a variational average-atom in a quantum plasma (VAAQP) was proposed and studied \cite{Blenski07a,Blenski07b,Piron11,Piron11b}. This showed that, building an atom-in-jellium model within a variational framework enables to set the jellium density from the thermodynamic equilibrium condition and to have the virial theorem fulfilled.

Formally, to treat the nuclei-electron plasma as a set of ions, we resort to a reasoning called a ``cluster'' decomposition. Let $O$ be a quantity which may be calculated for any set of $K$ nuclei, with spatial configuration $(\vec{R}_1 ... \vec{R}_K)$, including the empty set. We may then write (see \cite{Felderhof82} for more detail):
\begin{align}
O(\vec{R}_1 ... \vec{R}_K)
= O(\emptyset)
+\sum_{j=1}^{K} \Delta O_1(\vec{R}_j)
+\frac{1}{2}\sum_{j=1}^{K}\sum_{\substack{k=1\\k\neq j}}^{K} \Delta O_2 (\vec{R}_j,\vec{R}_k) +...
\label{eq_cluster_formal}
\end{align}
defining the $\Delta O_K$ terms recursively, so as to assure the equality for each value of $K$:
\begin{align}
&\Delta O_1(\vec{R})
=O(\vec{R})-O(\emptyset)
\\
&\Delta O_2(\vec{R}_1,\vec{R}_2)
=O(\vec{R}_1,\vec{R}_2)
-\Delta O_1(\vec{R}_1)
-\Delta O_1(\vec{R}_2)
+O(\emptyset)
\\
&\hspace{0.5cm}\vdots\nonumber
\end{align}
The quantity $O$ has a clustering property if the terms in Eq.~\eqref{eq_cluster_formal} exhibit a decreasing ordering, which makes the equation to be an convergent expansion. 

In the VAAQP model, we first assume that the electron density $n(\vec{R}_1 ... \vec{R}_{N_\text{i}};\vec{r})$ for a system of $N_\text{i}$ ion is correctly described limiting the cluster expansion to the zeroth and first order only:
\begin{align}
n(\vec{R}_1 ... \vec{R}_{N_\text{i}};\vec{r})
=n_0(\vec{r})+\sum_{j=1}^{N_\text{i}}\Delta n_1(\vec{R}_j;\vec{r})
=n_\text{e}+\sum_{j=1}^{N_\text{i}}q(|\vec{r}-\vec{R}_j|)
\label{eq_cluster_density}
\end{align}
where the zeroth order term is identified to the homogeneous jellium density, and the first order term correspond to the sum of spherically-symmetric clouds of displaced electrons, corresponding each to an ion in a jellium.
We also assume the first-order cluster expansion to hold for the free energy $F$ of the system. This leads us to write the free energy per ion as follows:
\begin{align}
\dot{F}(n_\text{i},T)=\dot{F}_0(n_\text{e};n_\text{i},T)+\Delta F_1\{n_\text{e},\underline{q};T\}
\label{eq_VAAQP_cluster_F}
\end{align}
Here, $\dot{F}_0=(f_\text{e}^\text{F}(n_\text{e},T)+f_\text{xc}(n_\text{e},T))/n_\text{i}$ is the free energy per ion of the uniform electron gas. We choose to treat $\Delta F_1$ using a density-functional formalism \cite{Hohenberg64,KohnSham65a,Mermin65} and decompose the $\Delta F_1$ as suggested by Kohn and Sham \cite{KohnSham65a}:
\begin{align}
\Delta F_1\{n_\text{e},\underline{q};T\}
=\Delta F_1^0\{n_\text{e},\underline{q};T\}
+\Delta F_1^\text{el}\{n_\text{e},\underline{q};T\}
+\Delta F_1^\text{xc}\{n_\text{e},\underline{q};T\}
\end{align}
$\Delta F_1^0$ corresponds to the kinetic and entropic contribution to the free energy of a system of independent electrons subject to an external potential $v_\text{trial}\left\{n_\text{e},\underline{q};r\right\}$ that yields the electron density $n_\text{e}+q(r)$, with the contribution from an homogeneous system of density $n_\text{e}$ subtracted.
\begin{align}
\Delta F_1^0\left\{n_\text{e},\underline{q};T\right\}
=\sum_\xi
\int d^3r \left\{
p_\text{F}(\varepsilon_\xi;n_\text{e},T)\left[\vphantom{\frac{1}{1}}
\left(\varepsilon_\xi-v_\text{trial}(r)-T s_\text{F}(\varepsilon_\xi;n_\text{e},T)\right)|\varphi_\xi(\vec{r})|^2
-\left(\varepsilon_\xi-T s_\text{F}(\varepsilon_\xi;n_\text{e},T)\right)|\varphi^0_\xi(\vec{r})|^2\right]
\right\}
\label{eq_VAAQP_dF10}
\end{align}
where the sum runs over both the discrete and continuum part of the spectrum, and where the $\{\varphi^0_\xi\}$ correspond to the plane waves and only contribute in the continuum part. 
$\Delta F_1^\text{xc}$ corresponds to the exchange and correlation contribution to the free energy, with the contribution from the homogeneous system subtracted, taken in the local density approximation:
\begin{align}
\Delta F_1^\text{xc}\{n_\text{e},\underline{q};T\}
=\int d^3r \left\{f_\text{xc}(n_\text{e}+q(r),T)-f_\text{xc}(n_\text{e},T)\right\}
\label{eq_VAAQP_dF1xc}
\end{align}

$\Delta F_1^\text{el}$ is the direct interaction term, in which we introduce the hypothesis of the WS cavity. We model the surrounding ions by a charge density $\ee n_\text{e}\theta(r-R_\text{WS})$, like in the AJCI model. This leads to:
\begin{align}
\Delta F_1^\text{el}=&
\int d^3r\left\{ -\frac{Z \left(n_\text{e}+q(r)-n_\text{e}\theta(r-R_\text{WS})\right) \ee^2}{r} 
\right.\nonumber\\&\left.
+\frac{\ee^2}{2}\int d^3r' \left\{ \frac{\left(n_\text{e}+q(r)-n_\text{e}\theta(r-R_\text{WS})\right)\left(n_\text{e}+q(r)-n_\text{e}\theta(r'-R_\text{WS})\right)}{|\vec{r}-\vec{r}'|}\right\} \right\}
\label{eq_VAAQP_dF1el}
\end{align}
Accordingly, the condition of neutrality in the whole space writes:
\begin{align}
Z=\int d^3r \left\{
n_\text{e}+q(r)-n_\text{e}\theta(r-R_\text{WS})
\right\}
\end{align}

Finally, the VAAQP model is based on the minimzation of the free energy with respect to the displaced-electron density $q(r)$ and jellium density $n_\text{e}$, while requiring the neutrality condition:
\begin{align}
\dot{F}_\text{eq}(N_\text{i},V,T)=&\underset{n_\text{e},\underline{q}}{\text{Min}}\,\dot{F}\left\{n_\text{e},\underline{q};n_\text{i},T\right\}
\text{ s.\,t. } \int d^3r \left\{
n_\text{e}+q(r)-n_\text{e}\theta(r-R_\text{WS})
\right\}=Z
\label{eq_minimiz_VAAQP}
\end{align}
This constrained minimization yields the following equations:
\begin{align}
&v_\text{trial}(r)=v_\text{el}(r)+\mu_\text{xc}(n_\text{e}+q(r))-\mu_\text{xc}(n_\text{e})\\
&v_\text{el}(r)=-\frac{Z \ee^2}{r}+\ee^2\int d^3r' \left\{ \frac{n_\text{e}+q(r)-n_\text{e}\theta(r'-R_\text{WS})}{|\vec{r}-\vec{r}'|}\right\}\\
&\int d^3r\left\{ v_\text{el}(r)\theta(r-R_\text{WS}) \right\}=0\label{eq_VAAQP_n0_condition}
\end{align}

Eq.~\eqref{eq_VAAQP_n0_condition} stems from the minimization condition with respect to the jellium density $n_\text{e}$ and allows to set its value. Thus, in the VAAQP model, the density of the uniform background is uniquely defined, it corresponds to the asymptotic electron density of each ion and is set by the thermodynamic equilibrium condition.

From the equilibrium free energy per ion, it is possible to obtain rigorously the other thermodynamic quantities, by calculating the relevant derivatives. For the pressure, the following formula is obtained:
\begin{align}
P_\text{thermo}=-f_\text{e}^\text{F}(n_\text{e},T)-f_\text{xc}(n_\text{e},T)
+n_\text{e}\mu^\text{F}_\text{e}(n_\text{e},T)+n_\text{e}\mu_\text{xc}(n_\text{e},T)+n_\text{e}v_\text{el}(R_\text{WS})
\label{eq_VAAQP_pressure}
\end{align}
The first four terms correspond to the pressure of an ideal Fermi gas of density $n_\text{e}$. Last term is related to the WS cavity. Moreover, it can be shown that the virial pressure leads to the same formula as Eq.~\eqref{eq_VAAQP_pressure}, meaning that the virial theorem is fulfilled in the VAAQP model.

In the contributions to the free energy expression of Eq.~\eqref{eq_VAAQP_cluster_F}, the ions are disregarded. For that reson, the thermodynamic quantities from the VAAQP model may be viewed as electron contributions, that may be supplemented by ion contributions. Adding a ion ideal-gas contribution to the model is straightforward. However, adding the results of a model of interacting ions is more problematic, because part of the ion-ion interactions are necessarily included in the VAAQP model through the WS-cavity hypothesis.

Like INFERNO, the VAAQP model allows the description of the ion shell structure, while the WS cavity assumed in the model enables the description of pressure ionization. Treating the perturbation of the density in the whole space, the model also accounts for the Friedel oscillations (see, for instance, \cite{AshcroftMermin}) of the displaced-electron density. The physical relevance of these oscillations in the case of ions in a plasma is rather unclear. However, accounting for these is essential to ensure fulfilment of the virial theorem.

In the VAAQP model, the potential range is not strictly limited to the WS radius but has a strong decay, due to the total screening of the nucleus in the whole space. In practice, the variational equation Eq.~\eqref{eq_VAAQP_n0_condition} most often constrains the atomic potential to take small values at the WS radius, of the order of the amplitude of Friedel oscillations. For that reason, the VAAQP model often yields results which closely agree with those of the INFERNO model, except in the low-temperature/high-density regime.

\begin{figure}[ht]
\centerline{
\includegraphics[width=8cm]{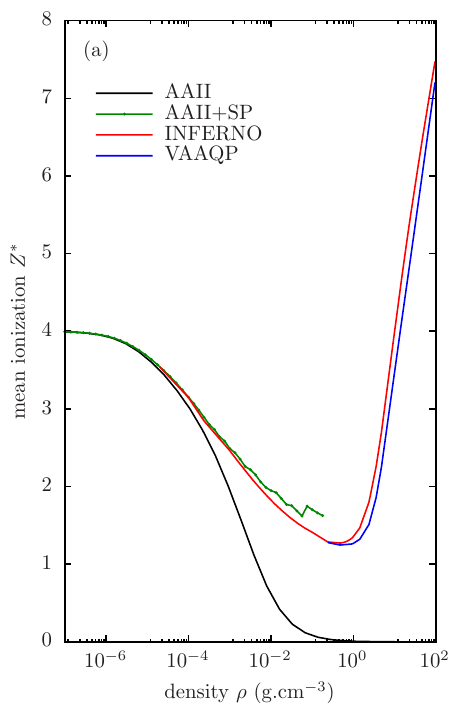}
\hspace{1cm}
\includegraphics[width=8cm]{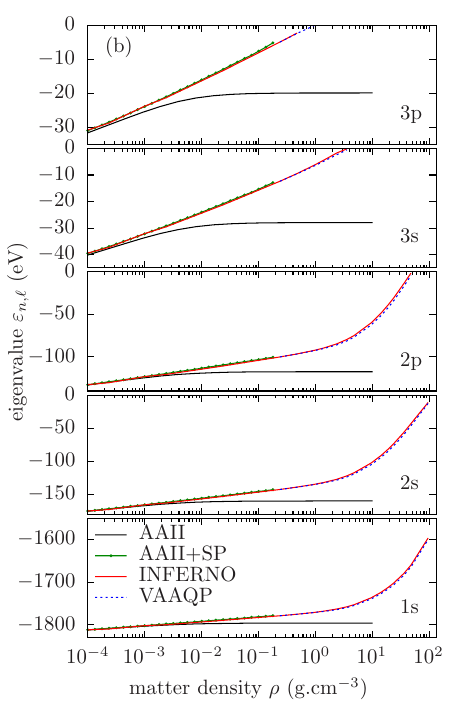}}
\caption{Mean ionization (a), and 1-electron eigenvalue (b) as a function of matter density, for silicon at 5-eV temperature. Comparisons between INFERNO, VAAQP, AAII and AAII with Stewart-Pyatt correction and suppression of orbitals
\label{fig_Si_ioniz_eigenvalues}}
\end{figure}

Figure~\ref{fig_Si_ioniz_eigenvalues} shows an example comparison of results from the isolated-ion, INFERNO, and VAAQP models in the case of silicon at 5-eV temperature. Both the mean ionization and the 1-electron energies are displayed.  As is seen from theses figures the energy correction, and consequently the mean ionization, are rather well estimated using the Stewart-Pyatt approach with the suppression of bound orbitals \cite{StewartPyatt66}, up to cases of significant pressure ionization (here, around 0.1 g.cm$^{-3}$). INFERNO and VAAQP models agree well in this regime. In the region of strong pressure ionization, results from VAAQP depart significantly from those of INFERNO. Accordingly, thermodynamic consistency of the INFERNO results is problematic in this region. However, differences on 1-electron energies are less pronounced.

The VAAQP model was applied to equation-of-state calculations \cite{Piron11,Piron11b}, but was also used in the calculation of radiative properties \cite{Piron13,Piron18}.

Finally, atom-in-jellium models seem a better starting point for improving the modeling of ion-ion correlations than ion-cell models, since they account for the ion surrounding in the whole space.

\subsection{Going beyond the cavity hypothesis, the difficulty of dense-plasmas atomic modeling}
All  models of pressure-ionized plasma which are described in the previous paragraphs focus on the description of the electronic structure of a particular ion, postulating that the surrounding ions will either restrict its spatial extension to a WS cell, or will interact with it as a spread-out uniform medium, not entering the WS sphere. In these models, a key function is the inhomogeneous density $n(r)$ of the electron cloud associated to the ion. Of course such spatial inhomogeneity of the electron density around a nucleus is in fact refering to correlations among the positions of nuclei and electrons.

For an $M$-component classical fluid of particles interacting through pair potentials, obtaining the $M(M+1)/2$ pair distribution functions gives access to all statistical averages (see, for instance \cite{LandauStatisticalPhysics}). The link relating the pair distribution functions to densities of fictitious systems with a particular particle placed at the origin is the Percus picture \cite{Percus64}. With such a link, one can use models of 1-particle densities to address pair distribution functions.

However, to describe the electronic structure of an ion one has to resort to quantum mechanics. It turns out that Percus' method is not applicable in the quantum-mechanical context. This may be seen as a consequence of the impossibility of separating the kinetic part of the partition function from its configurational (or interaction) part. For that reason, a practical approach to the correlation functions in quantum mechanics remains a long-standing, open problem (see, for instance \cite{Percus05}).

In order to circumvent this problem, an idea is to keep the framework of classical statistical mechanics and include an approximate accounting for the quantum behavior in the interaction potentials \cite{Lado67}, or resorting to an effective temperature \cite{Dharma-wardana00,Perrot00}. Approaches of this kind eventually led to the classical-mapping approach to quantum systems \cite{Dufty13,Dutta13}.

Another kind of approach aim at extending Percus' trick (see, for instance \cite{Chihara78}). Among the goals of such effort is the quantum generalization of the hypernetted chain model, that may be derived in the classical framework using the Percus picture and the classical DFT (see, for instance \cite{HansenMcDonald}). As was pointed out by Chihara \cite{Chihara91}, if one assumes that nuclei behave as classical particles, it is possible to apply partially the Percus trick. This requires to supplement the model with assumptions about some of the correlation functions. This approach led to the ``quantum hypernetted chain'' (QHNC) model. 

However, the DFT-based reasoning for deriving the HNC equations resorts to a functional Taylor expansion around a reference homogeneous medium at given density. Even if this derivation leads to the HNC equations, it does not give access to the HNC free energy of the homogeneous system (see, for instance \cite{DeWitt88} and \cite{Piron19a} appendix). This is an issue for the modeling of the thermodynamic equilibrium of a plasma, which includes the determination of the free-electron density. For that reason, the QHNC approach was first used in the modeling of electron in metals \cite{Anta00}, considering given free-electron densities. The QHNC was also applied to modeling of dense plasmas \cite{Starrett12,Starrett13,Starrett14,Chihara16}, in order to account for the ion-ion correlations. However, it requires in this case an additional assumption to set the density of free electrons.

\subsection{VAMPIRES model}
A way towards an improved modeling of pressure-ionized plasma is to couple an atomic model of plasma to a classical model of fluid through its interaction potentials, without resorting to the point-like-ion hypothesis to split the problem. 

In such a model, one should account for the impact of the ion-fluid structure on the electronic structure of ions, but also for the effect of the electronic structure on the  interaction potentials in the ion fluid. The interaction potentials are then to be determined self-consistently. Ideally, they should be seen as thermodynamic averages, obtained from the minimization of the total free energy. 

For these reasons, an essential building block for such a model is a generalized free-energy functional related to a statistical model of a simple fluid, for an arbitrary interaction potential. A free-energy functional of the interaction potential was available since the early 60's for the HNC model \cite{MoritaHiroike60,Lado73}. The equivalent for the Debye-H\"uckel model was developped more recently \cite{Piron16,Blenski17,Piron19a}.

Moreover, such free energy functionals can be formulated as minima of the related generalized free energy functionals of the pair distribution function. This offers an elegant way of deriving the integral equations of the corresponding fluid model, through a minimization with respect to the pair distribution function.

A preliminary work on a model accounting for both the bound electrons of an ion and the ion fluid structure was described in \cite{Piron19b}. In this model, continuum electrons are excluded from the ion electronic structure, as in an isolated-ion model, and participate in a species of a classical fluid. This classical fluid may be treated either through the Debye-H\"{u}ckel model (thus avoiding the Coulomb collapse), or neglecting the polarization of the electrons as in an OCP. Despite the obvious limitation in its applicability, due to its crude treatment of continuum electrons, this model formally introduces the screening of the effective potential in the electronic structure. In the DH case, when bound electrons are localized in a small region compared to the Debye length, this model yields the point-like DH correction of Eq.~\eqref{eq_DH_AA_correction}.

The variational atomic model of plasma with ion radial correlations and electronic structure (VAMPIRES) \cite{BlenskiPiron23} is both an atom-in-jellium model of the ion electronic structure and a statistical model of ion fluid. In this model, the continuum electrons are treated quantum-mechanically, as a part of the electronic structure partially shared among ions. This model stems from the minimization of an approximate free energy and it was shown to be fulfilling the virial theorem. 

Let us consider the free energy of $N_\text{i}$ nuclei of atomic number $Z$ and $N_\text{i} Z$ electrons, in a large volume $V$ and at a fixed temperature $T$. The nuclei are approximated by undistinguishable classical particles, which allows us to write (see, \cite{Kirkwood33,Zwanzig57}, and \cite{BlenskiPiron23} for the present generalized form):
\begin{align}
&F_\text{eq}(N_\text{i},V,T)
\nonumber\\
&=\underset{\underline{w}}{\text{Min}}\,\iint_V \frac{d^3R_1...d^3P_{N_\text{i}}}{N_\nu !h^{3N_\text{i}}}\left\{
w(\vec{R}_1...\vec{P}_{N_\text{i}})
\left({\sum_{j=1}^{N_\text{i}} \frac{P_j^2}{2m_\text{i}}}
+F_\text{eq}^\text{e}(\vec{R}_1...\vec{R}_{N_\text{i}};N_\text{i},V,T)
+{\frac{1}{\beta}\log\left(w(\vec{R}_1...\vec{P}_{N_\text{i}})\right)}
\right)
\right\}
\nonumber\\
&\text{\hspace{1cm}s. t. } \iint_V \frac{d^3R_1...d^3P_{N_\text{i}}}{N_\text{i}!h^{3N_\text{i}}}\left\{w(\vec{R}_1...\vec{P}_{N_\text{i}})\right\}=1
\label{eq_VAMPIRES_general_minimiz}
\end{align}
where $w(\vec{R}_1...\vec{P}_{N_\text{i}})$ denotes the probability distribution of the nuclei classical many-body states $(\vec{R}_1...\vec{P}_{N_\text{i}})$, and where $F_\text{eq}^\text{e}$ is the equilibrium free energy of a system of electron with a fixed configuration $(\vec{R}_1...\vec{R}_{N_\text{i}})$ of the nuclei, plus the nucleus-nucleus interaction energy. The constraint simply enforce the correct normalization of the probability.

Electrons are modeled quantum-mechanically, using a finite-temperature density-functional formalism \cite{Hohenberg64,KohnSham65a,Mermin65}. That is we obtain $F_\text{eq}^\text{e}$ from the following minimization:
\begin{align}
F_\text{eq}^\text{e}(\vec{R}_1...\vec{R}_{N_\text{i}};N_\text{i},V,T)=&\underset{\underline{n}}{\text{Min}}\,\left[{F^0\left\{\underline{n};V,T\right\}}
+W_\text{direct}\left\{\underline{n};\vec{R}_1...\vec{R}_{N_\text{i}};N_\text{i}\right\}
+{F^\text{xc}\left\{\underline{n};V,T\right\}}\right]
\nonumber\\
&~\text{ s. t. } \int_V d^3r\left\{n(\vec{r})\right\}=ZN_\text{i}
\label{eq_VAMPIRES_DFT_minimiz}
\end{align}
$n(\vec{r})$ is the electron density, $F^0$ denotes the kinetic-entropic contribution to the free energy of a non-interacting electrons gas of density $n(\vec{r})$, $W_\text{direct}$ denotes the total direct-interaction energy, which includes the nucleus-nucleus contribution. $F^\text{xc}$ is the contribution of exchange and correlations. The constraint corresponds to the neutrality condition of the nuclei-electron system.

Like in the VAAQP model, we assume that the equilibrium electron density $n(\vec{R}_1 ... \vec{R}_{N_\text{i}};\vec{r})$ for a system of $N_\text{i}$ nuclei is correctly described using a first-order cluster expansion (see Eq.~\eqref{eq_cluster_density}). This leads us to the following Ansatz for the electron density:
\begin{align}
n(\vec{R}_1...\vec{R}_{N_\text{i}};\vec{r})
\approx n_\text{e}+\sum_{j=1}^{N_\text{i}} q(|\vec{r}-\vec{R}_j|)
\label{eq_ansatz_VAMPIRES}
\end{align}
The system is seen as a set of ions, that is: a nucleus with its spherical cloud of displaced electrons, sharing a common uniform background of free electrons. The minimization with respect to the electron density $n(r)$ is thus performed within a particular class of functions, and consists in a minimization with respect to the two parameters of the Ansatz, namely $n_\text{e}$ and the function $q(r)$.

The neutrality condition of Eq.~\eqref{eq_VAMPIRES_DFT_minimiz} can be rewritten using Eq.~\eqref{eq_ansatz_VAMPIRES}, as:
\begin{align}
\frac{n_\text{e}}{n_\text{i}}+\int_V d^3r\left\{q(\vec{r})\right\}=Z
\label{eq_VAMPIRES_neutrality}
\end{align}

Using Eq.~\eqref{eq_ansatz_VAMPIRES}, $W_\text{direct}$ can be recast as:
\begin{align}
W_\text{direct}
={\frac{1}{2}\sum_{i=1}^{N_\text{i}}\sum_{\substack{j=1\\j\neq i}}^{N_\text{i}}v_\text{ii}\left\{\underline{q};|\vec{R}_i-\vec{R}_j|\right\}}
+{N_\text{i}\,W_\text{intra}\left\{\underline{q};V\right\}}
+{N_\text{i}\,W_\text{bg}\left\{\underline{q},n_\text{e};V\right\}}
\end{align}
with the definitions:
\begin{align}
&v_\text{ii}\left\{\underline{q};R,V\right\}
=
\frac{Z^2\ee^2}{R}
-2Z\ee^2\int_V d^3r \left\{ \frac{q(r)}{|\vec{r}-\vec{R}|} \right\}
+\ee^2\int_V d^3r d^3r' \left\{ \frac{q(r)q(r')}{|\vec{r}-\vec{r}'+\vec{R}|} \right\}
\nonumber\\
&W_\text{intra}\left\{\underline{q};V\right\}=
-Z\ee^2\int_V d^3r \left\{ \frac{q(r)}{r} \right\}
+\frac{\ee^2}{2}\int_V d^3r d^3r' \left\{ \frac{q(r)q(r')}{|\vec{r}-\vec{r}'|} \right\}
\nonumber\\
&W_\text{bg}\left\{\underline{q},n_\text{e};V\right\}=
n_\text{e}\ee^2\int_V d^3rd^3r'\left\{ \frac{q(r)}{|\vec{r}-\vec{r}'|}\right\}
+\ee^2\left( \frac{n_\text{e}^2}{2n_\text{i}}-n_\text{e}Z \right)\int_V d^3r \left\{ \frac{1}{r} \right\}
\end{align}
$v_\text{ii}$ plays the role of an ion-ion interaction potential, $W_\text{intra}$ corresponds to an intra-ion interaction energy, and $W_\text{bg}$ gathers all terms related to interactions with the electron homogeneous background. Contrary to what was done in VAAQP, the electrostatic interaction term pertaining to an ion is not the object of a specific hypothesis like Eq.~\eqref{eq_VAAQP_dF1el}. In the VAMPIRES model, the interaction terms just follows from the cluster expansion of the electron density and the statistical treatment of the ion fluid.

Electron terms $F^0$ and $F^\text{xc}$ are approximated using a first-order cluster expansion, as in VAAQP: 
\begin{align}
F^{\bullet}&\left\{\underline{n}(\vec{r})=n_\text{e}+\sum_{i=1}^{N_\text{i}}q(|\vec{r}-\vec{R}_i|);V,T\right\}
= F^{\bullet}\left\{\underline{n}(\vec{r})=n_\text{e};V,T\right\}
+\sum_{i=1}^{N_\text{i}} \Delta F_1^{\bullet}\left\{\underline{q},n_\text{e},\vec{R}_i;V,T\right\}
\end{align}
\begin{align}
\Delta F_1^{\bullet}\left\{\underline{q},n_\text{e},\vec{R};V,T\right\}
=&F^{\bullet}\left\{\underline{n}(\vec{r})=n_\text{e}+q(|\vec{r}-\vec{R}|);V,T\right\}-F^{\bullet}\left\{\underline{n}(\vec{r})=n_\text{e};V,T\right\}
=\Delta F_1^{\bullet}\left\{\underline{q},n_\text{e};V,T\right\}
\end{align}
where the $^{\bullet}$ symbol is to be replaced by either the $^0$ or the $^\text{xc}$ label.
$\Delta F_1^0\left\{\underline{q},n_\text{e};V,T\right\}$ is the kinetic and entropic contribution to the free energy of non-interacting electrons in a trial potential $v_\text{trial}\left\{\underline{q},n_\text{e};r;T\right\}$, that yields the electron density $n(r)=n_\text{e}+q(r)$, minus the contribution of the homogeneous background (see Eq.~\eqref{eq_VAAQP_dF10}).
$\Delta F_1^\text{xc}\left\{\underline{q},n_\text{e};V,T\right\}$ is the exchange-correlation contribution to the free energy of a system of electrons having density $n(r)=n_\text{e}+q(r)$, minus the contribution of the homogeneous background (see Eq.~\eqref{eq_VAAQP_dF1xc}).

At this point, the minimization of Eq.~\eqref{eq_VAMPIRES_general_minimiz} becomes
\begin{align}
F_\text{eq}(N_\text{i},V,T)=&
\underset{\underline{q},n_\text{e}}{\text{Min}}
\left[F^0\left\{n_\text{e};V,T\right\}
+F^\text{xc}\left\{n_\text{e};V,T\right\}
\right.\nonumber\\&\left.
+N_\text{i}\left( 
\Delta F_1^0\left\{\underline{q},n_\text{e};V,T\right\}
+\Delta F_1^\text{xc}\left\{\underline{q},n_\text{e};V,T\right\}
+W_\text{intra}\left\{\underline{q},V\right\}
+W_\text{bg}(\underline{q},n_\text{e};V)
\right)
\right.\nonumber\\&\left.
+F_\text{eq}^\text{i}
\left\{
\underline{v}(R)=v_\text{ii}\left\{\underline{q};R,V\right\}
;N_\text{i},V,T
\right\}
\right]
\nonumber\\
&\text{s.\,t. }\frac{n_\text{e}}{n_\text{i}}+\int_V d^3r\left\{q(r)\right\}=Z
\label{eq_free_energy_one-ion}
\end{align}
where $F_\text{eq}^\text{i}\left\{\underline{v};N_\text{i},V,T\right\}$ gathers the nuclei kinetic energy, entropy, and ion-ion interaction terms, forming the free energy of a one-component classical fluid of ions, interacting through the potential $v_\text{ii}$:
\begin{align}
F_\text{eq}^\text{i}\left\{\underline{v};N_\text{i},V,T\right\}
=&\underset{\underline{w}}{\text{Min}}
\int_V \frac{d^3R_1...d^3P_{N_\text{i}}}{N_\text{i}!h^{3N_\text{i}}}\left\{
w(\vec{R}_1...\vec{P}_{N_\text{i}})
\left(
\sum_{j=1}^{N_\text{i}} \frac{P_j^2}{2m_\text{i}}
+\frac{1}{2}\sum_{i=1}^{N_\text{i}}\sum_{\substack{j=1\\j\neq i}}^{N_\text{i}}v(|\vec{R}_i-\vec{R}_j|)
+\frac{1}{\beta}\log\left(w(\vec{R}_1...\vec{P}_{N_\text{i}})\right)
\right)
\right\}
\nonumber\\
&\text{s.\,t. }\int_V \frac{d^3R_1...d^3P_{N_\text{i}}}{N_\text{i}!h^{3N_\text{i}}}\left\{w(\vec{R}_1...\vec{P}_{N_\text{i}})\right\}=1\\
\equiv&\underset{\underline{w}}{\text{Min}}\,
F^\text{i}\left\{\underline{w},\underline{v};N_\text{i},V,T\right\}
\text{ s.\,t. }\int_V \frac{d^3R_1...d^3P_{N_\text{i}}}{N_\text{i}!h^{3N_\text{i}}}\left\{w(\vec{R}_1...\vec{P}_{N_\text{i}})\right\}=1
\label{eq_general_classical_fluid}
\end{align}

In the thermodynamic limit, the free energy per ion $\dot{F}_\text{i}={F}_\text{i}/N_\text{i}$ of such a system has a logarithmic divergence, because $v_\text{ii}$ has a Coulomb tail. However, as in a usual OCP model, this divergence is cancelled by an opposite-sign divergence in $W_\text{bg}$. We therefore group these terms together, which renormalizes the free energy. We use either the HNC or the DH model to approximate the resulting divergence-free ion-fluid free energy per ion, as a functional of the interaction potential $v(r)$.
\begin{align}
\dot{F}^\text{i}_\text{eq}+{W_\text{bg}}
\approx \dot{F}_\text{id, i}(n_\text{i},T)
+\dot{F}_\text{ex, eq}^\text{approx}\left\{\underline{v};n_\text{i},T\right\}
\end{align}
$\dot{F}_\text{ex, eq}^\text{approx}$ being either the HNC or the DH excess free-energy per ion.
Such approximate equilibrium free energy may be written as the minimum of a generalized free-energy functional of the radial correlation function $h(r)=g(r)-1$:
\begin{align}
\dot{F}_\text{ex, eq}^\text{approx}\left\{\underline{v};n_\text{i},T\right\}
=\underset{\underline{h}}{\text{Min}}\,\dot{F}_\text{ex}^\text{approx}\left\{\underline{v},\underline{h};n_\text{i},T\right\}
\end{align}
with the minimum occuring for $h(r)$ fulfilling the equations of the approximate model, either HNC or DH.
In the DH case, we have \cite{Piron19a}:
\begin{align}
\dot{F}_\text{ex}^{\text{DH}}
=&
\frac{n_\text{i}}{2\beta}\int d^3r \left\{h(r)\beta v(r) \right\}
+\frac{1}{2\beta n_\text{i}}\int \frac{d^3k}{(2\pi)^3}
\left\{ n_\text{i} h_{k}-\log(1+n_\text{i} h_{k}) \right\}
\end{align}
whereas in the case of HNC, we have \cite{MoritaHiroike60,Lado73}:
\begin{align}
\dot{F}_\text{ex}^\text{HNC}
=\dot{F}_\text{ex}^{\text{DH}}+&\frac{n_\text{i}}{2\beta}
\int d^3r\left\{\vphantom{\frac{(r)^2}{2}}
(h(r)+1)\log\left(h(r)+1\right)
-h(r)-\frac{h(r)^2}{2}\right\}
\end{align}

Finally, the VAMPIRES model is based on the minimization of the following approximate free energy per ion $\dot{F}\{\underline{h},\underline{q},n_\text{e}\}$, with the neutrality constraint:
\begin{align}
\dot{F}\{\underline{h},\underline{q},n_\text{e}\}
=&
\dot{F}_\text{id, i}(n_\text{i},T)
+\dot{F}_\text{ex}^\text{i\,approx}\left\{\underline{h},\underline{v}(R)
=v_\text{ii}\left\{\underline{q},R\right\};n_\text{i},T\right\}
\nonumber\\&
+\frac{f^\text{F}_\text{e}(n_\text{e};T)}{n_\text{i}}
+\frac{f_\text{xc}(n_\text{e};T)}{n_\text{i}}
+\Delta F_1^0\left\{\underline{q},n_\text{e};T\right\}+\Delta F_1^\text{xc}\left\{\underline{q},n_\text{e};T\right\}+W_\text{intra}\left\{\underline{q}\right\}
\\
\dot{F}_\text{eq}(N_\text{i},T)
=&\underset{\underline{h},\underline{q},n_\text{e}}{\text{Min}}\,\dot{F}\{\underline{h},\underline{q},n_\text{e}\}
~\text{ s. t. }\frac{n_\text{e}}{n_\text{i}}+\int d^3r\left\{q(r)\right\}=Z
\end{align}

The minimisation with respect to $h(r)$ leads to the fluid integral equations, that is: the Ornstein-Zernike relation with the closure relation corresponding to the chosen approximate model:
\begin{align}
&h(r)=c(r)
+n_\text{i}\int d^3r'\left\{
c(|\vec{r}'-\vec{r}|)h(r') 
\right\}
\\
&c(r)=-\beta v_\text{ii}(r)-\log(h(r)+1)+h(r)&\text{(HNC)}
\\
&c(r)=-\beta v_\text{ii}(r)&\text{(DH)}
\end{align}

The minimization with respect to $q(r)$ includes that on $n_\text{e}$, which is expressed as a functional of $q(r)$ due to the neutrality constraint. It yields:
\begin{align}
0=&-v_\text{trial}\left\{\underline{q},n_\text{e};r;T\right\}
+\mu_\text{xc}\left(n_\text{e}+q(r),T\right)-\mu_\text{xc}\left(n_\text{e},T\right)+{v}_\text{el}(r)
\nonumber\\
&-n_\text{i}\int d^3r' \left\{
-v_\text{trial}\left\{\underline{q},n_\text{e};r';T\right\}
\right.\left.
+\mu_\text{xc}\left(n_\text{e}+q(r'),T\right)-\mu_\text{xc}\left(n_\text{e},T\right) \right\}\label{eq_VAMPIRES_SCF_intermediary}
\end{align}
where we have defined:
\begin{align}
&v_\text{el}\left\{\underline{h},\underline{q},n_\text{e};r;T\right\}
\equiv v_\text{intra}\left\{\underline{q};r\right\}
+n_\text{i}\int d^3r' \left\{h(r')v_\text{intra}\left\{\underline{q};|\vec{r}'-\vec{r}|\right\}\right\}
\label{eq_VAMPIRES_vel}
\\
&v_\text{intra}\left\{\underline{q};r\right\}
=\frac{-Z\ee^2}{r}+\ee^2\int d^3r'\left\{\frac{q(r')}{|\vec{r}-\vec{r}'|}\right\}
\end{align}
In order to solve Eq.~\eqref{eq_VAMPIRES_SCF_intermediary}, we define the distribution $\tilde{v}_\text{el}$ such that: 
\begin{align}
v_\text{el}(r) = \tilde{v}_\text{el}\left\{\underline{v}_\text{el},r\right\}	- n_{\nu}\int d^3r'\left\{\tilde{v}_\text{el}\left\{\underline{v}_\text{el},r'\right\}\right\}
\label{eq_VAMPIRES_eq_vel_veltilde}
\end{align}
We thus obtain from Eq.~\eqref{eq_VAMPIRES_SCF_intermediary} the following electron self-consistent equation:
\begin{align}
v_\text{trial}\left\{\underline{q},n_\text{e};r;T\right\}=\tilde{v}_{\text{el}}(r)
+\mu_\text{xc}\left(n_\text{e}+q(r),T\right)-\mu_\text{xc}\left(n_\text{e},T\right)
\end{align}
From Eq.~\eqref{eq_VAMPIRES_eq_vel_veltilde}, $\tilde{v}_\text{el}$ may be expressed in the Fourier space as:
\begin{align}
&\tilde{v}_{\text{el},k} = 
\begin{cases}			
v_{\text{el},k}=- \frac{4\pi e^{2}}{k^{2}}(Z-q_k)(1+n_{\nu}h_k) & \text{if } 	\vec{k}\neq 0\\
0 					  &	\text{if }  \vec{k}= 0
\end{cases}	
\end{align}
The difference between $\tilde{v}_\text{el}$ and ${v}_\text{el}$ only impacts on integrals of product of $\tilde{v}_\text{el}$ with a function that is not regular at $\vec{k}=0$ in the Fourier space. For instance, we have:
\begin{align}
\int d^3r\left\{\tilde{v}_\text{el}(r)\right\}=0
\neq
\int d^3r\left\{{v}_\text{el}(r)\right\}=-\frac{\beta n_\text{e}}{n_\nu}
\end{align}
where the last equality may be shown from the equations of the model.

Thermodynamic quantities are rigorously derived from the equilibrium free energy. Especially, the pressure is given by the following expression:
\begin{align}
P_\text{thermo}(n_\text{i},T)=&n_\text{i} k_\text{B} T
+\left.n_\text{i}^2
\frac{\partial {A}^\text{i\,approx}}{\partial n_\text{i}}
\right|_\text{eq}
+
n_\text{e} \big(\mu(n_\text{e},T)+\mu_\text{xc}(n_\text{e},T)\big) - f_0(n_\text{e},T) - f_\text{xc}(n_\text{e},T)
\end{align}
In this formula, the first two terms correspond to the pressure of the ion fluid (ideal-gas and excess contributions), while the next four terms correspond to the pressure of the uniform electron gas, as in the VAAQP model. This means that displaced electrons only contribute to the pressure through the ion-fluid excess term. From the expression of the virial pressure, it can also be shown that the virial theorem is fulfilled in the VAMPIRES model.

\begin{figure}[ht]
\centerline{
\includegraphics[width=8cm]{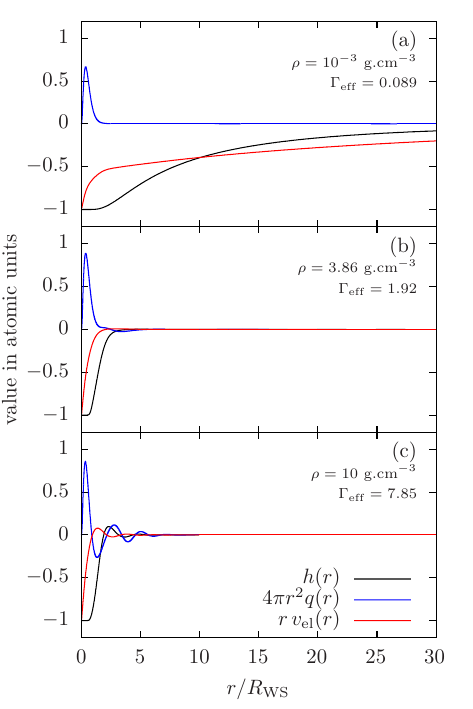}
\hspace{1cm}
\includegraphics[width=8cm]{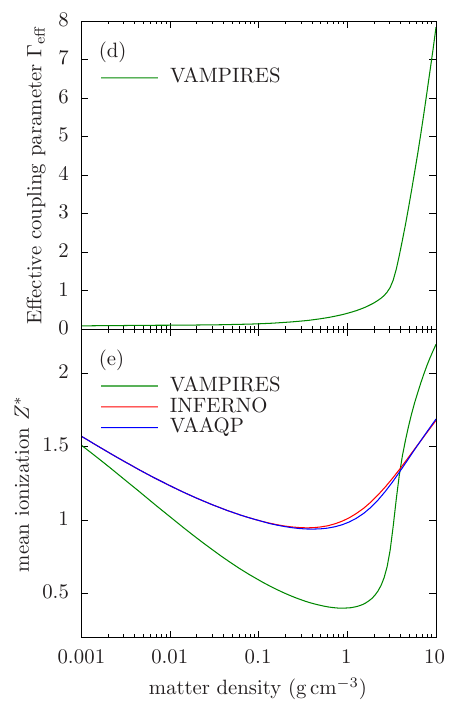}
}
\caption{Results from the VAMPIRES model for lithium at 10 eV temperature. Pair correlation function $h(r)$, electron density $4\pi q(r)/Z$, and electrostatic potential $r\,v_\text{el}(r)/Z$, for various matter densities (a,b,c). Mean ionization $Z^*$ (e) and effective coupling parameter $\Gamma_\text{eff}$ (d) as functions of the matter density.
\label{fig_Li_VAMPIRES}}
\end{figure}

Figure~\ref{fig_Li_VAMPIRES} presents results from the VAMPIRES model for lithium at 10 eV temperature. First, one sees from Fig.~\ref{fig_Li_VAMPIRES}e, which displays the mean ionzation as a function of density, that the accounting for ion-ion correlations in the model efficiently yields the qualitative behavior of pressure ionization. 

In order to quantify ion-ion coupling in this model, the usual coupling parameter $\Gamma=\beta Z^{*\,2}\ee^2/R_\text{WS}$ is not relevant. The ion-ion potential is not purely Coulombic, and ion charge $Z^*$ corresponds to an asymptotic limit, in general not relevant to the WS radius. Consequently, we use an effective coupling parameter $\Gamma_\text{eff}=-\beta \dot{U}_\text{ex}^\text{approx}$, which really corresponds to the ratio of the ion-fluid interaction energy to the thermal energy. 

Figure~\ref{fig_Li_VAMPIRES}a,b,c present the ion-ion pair correlation function $h(r)$, electron-cloud density $q(r)$ and ion effective electrostatic potential $v_\text{el}(r)$, for three values of matter densities corresponding to weak, moderate and strong coupling, respectively. In each case, close to the central nucleu, one sees a sharp peak in the electron linear density, which corresponds to the bound electronic structure of the ion. 

For a weakly-coupled plasma (case of Fig.~\ref{fig_Li_VAMPIRES}a), one can see that the range of the potential $v_\text{el}$ felt by the electrons extends way beyond the WS sphere. $v_\text{el}$ variations may be decomposed in two regions. Close to the nucleus, the steep variation is related to the ``internal'' screening by the bound electronic structure. Far from the nucleus, the longer-range decay is related to both a tail of weakly displaced electrons and the DH-like decay of the ion-ion correlation function.

Besides, one can see on Fig.~\ref{fig_Li_VAMPIRES}e that mean ionization in these cases is lower than in INFERNO or VAAQP. However, in VAMPIRES, some of the electrons that do not participate in the background density $n_\text{e}$, which defines $Z^*$, may in fact be weakly displaced and play a role similar to the background electrons in an observable quantity. Especially, these electrons may interact significantly with the surrounding ions. 

For a moderately-coupled plasma (case of Fig.~\ref{fig_Li_VAMPIRES}b), the ion-ion pair correlation function has the shape of a cavity, ressembling to the WS cavity assumed in VAAQP. In such situations, results from the VAMPIRES model are indeed close to those of VAAQP, or INFERNO. One may check on the figure that the range of $v_\text{el}$ is close to $R_\text{WS}$.

For a strongly-coupled plasma (case of Fig.~\ref{fig_Li_VAMPIRES}c), the ion-ion pair correlation function exhibits oscillations beyond the WS radius, which are typical of liquid-like behavior. According to the model equation Eq.~\eqref{eq_VAMPIRES_vel}, the correlation peaks of $h(r)$ draw some electrons. They are also ``dressed'' with the ion electron cloud density. This generates repulsive features between the central nucleus and the first correlation peak and between the correlations peaks, because of the potential overlap of electron clouds. Consistently, electrons are displaced away from these regions of potential overlap. As a consequence, $v_\text{el}$ has a zero inside the WS sphere, and its effective range is thus shorter than $R_\text{WS}$.

In this model, it seems that the pressure ionization phenomenon goes along with the switching to the liquid-like regime. This is illustrated on Fig.~\ref{fig_Li_VAMPIRES}d and e. The increase of the mean ionization is connected to a sharp increase of the coupling parameter. Across the pressure ionization edge, the plasma switches from a moderate-coupling to a strong-coupling regime, with the related feedback on the range of $v_\text{el}$. In addition to decreasing the value of $R_\text{WS}$, the range of $v_\text{el}$ switch from longer than $R_\text{WS}$ to shorter than $R_\text{WS}$. This explains why pressure ionization leads to a steeper increase of the mean ionization in this model than in VAAQP or INFERNO.

In the VAMPIRES model, the pressure-ionization phenomenon, as well as the switching from the Debye-Huckel-scale to the WS-scale decay stem from a first-principle accounting for the structure of the ion fluid. At the same time, at each thermodynamic condition, the ionization state of the plasma is obtained from the condition of thermodynamic equilibrium.

Among the known modeling issues for such a model is that the electrons of an ion feel the surrounding ions through their average distribution, given by the pair correlation function. Among the outcomes of such modeling effort could be that this static picture breaks down at some thermodynamic conditions.

Work on the VAMPIRES model is still in progress, and applicational studies are still to be performed. Therefore the following sections will only focus on applications of cavity-based models, which may however give relevant information.

\section{Radiative processes and photoabsorption in dense plasmas}
From the electrodynamics of continuous media, we can relate the photoabsorption cross-section per ion $\sigma_\text{abs}$ to the plasma dielectric function $\epsilon_\omega$ as follows \cite{LandauElectrodynamicsContinuousMedia}: 
\begin{align}
&\sigma_\text{abs}(\omega) = 
\frac{\omega \text{Re}(n_\omega^\text{ref})}
{n_\text{i}\epsilon_0 c}
\text{Im}(\epsilon_\omega)\label{eq_sigma_abs}
\\
&\text{Re}(n_\omega^\text{ref})
=\left(\frac{\text{Re}(\epsilon_\omega)+|\epsilon_\omega|}{2\epsilon_0}\right)^{1/2}\label{eq_real_nref}
\end{align}
where $n_\omega^\text{ref}$ correspond to the refraction index of the plasma.

One then decompose the dielectric function of the plasma into contributions from the various ions. Of course, such a decomposition depend on the considered atomic model. In the case of an ideal plasma of isolated ions, the response of the plasma is directly the sum of the responses of each isolated ion. In the case of an ion-cell model, one sees the the response of the plasma as the sum of the responses of each ion cell. For an atom-in-jellium model such as VAAQP, the cluster expansion \cite{Felderhof82}, which is used for the free energy, is extended to the plasma susceptibility \cite{Blenski92,Blenski94,Felderhof95a}.

In the dipole approximation, valid for wavelengths large compared to the typical atomic radius, we may write, limiting ourselves to the average-atom framework:
\begin{align}
\text{Im}(\epsilon_\omega)= \frac{n_\text{i} \ee^2}{\hbar}\, \text{Im}\int d^3r d^3r'\left\{z z'\mathcal{D}_\omega^\text{R}(\vec{r},\vec{r}')\right\}
\end{align}
where $\mathcal{D}_\omega^{R}$ is the atomic retarded susceptibility of the average atom, which can be written, at finite temperature as (see, for instance \cite{FetterWalecka}):
\begin{align}
\mathcal{D}_\omega^\text{R}(\vec{r},\vec{r}')
=-\frac{i}{\hbar}\int_{-\infty}^{+\infty} d\tau \left\{ 
\text{Tr}\left(\hat{\rho}\left[\hat{n}_{\vec{r}}^H(\tau),\hat{n}_{\vec{r}'}^H(0)\right]\right)\theta(\tau)e^{i\omega\tau}
\right\}
\end{align}
In the framework of the time-dependent density-functional theory
\cite{Stott80b,Zangwill80,Runge84,Dhara87,Gosh88}, one can properly relate the density susceptibility to the response of the electron density to a frequency-dependent external potential:
\begin{align}
\delta n_\omega(\vec{r})
=
\int d^3r'\left\{\mathcal{D}_\omega^\text{R}(\vec{r},\vec{r}')
\delta v_{\text{ext},\omega}(\vec{r}')\right\}
\end{align}
where $\delta v_{\text{ext},\omega}(\vec{r}')$ is the frequency-dependent perturbation of the potential, and $\delta n_\omega(\vec{r})$ is the resulting perturbation in the frequency-dependent density.

\subsection{Independent particle approximation and the effect of screening}
The simplest approximation for the retarded susceptibility is to use the retarded susceptibility of a system of independent particles. The latter susceptibility is obtained directly from the dynamic perturbation theory, which yields:
\begin{align}
&\mathcal{D}_\omega^\text{R}(\vec{r},\vec{r}') \approx \mathcal{D}_\omega^{\text{R},0}(\vec{r},\vec{r}')=\sum_{\xi,\zeta}\left(p_\text{F}(\mu,T,\varepsilon_\xi)-p_\text{F}(\varepsilon_\zeta)\right)
\frac{\varphi_\xi^*(\vec{r})\varphi_\xi(\vec{r}')\varphi_\zeta(\vec{r})\varphi_\zeta^*(\vec{r}')}
{\varepsilon_\zeta-\varepsilon_\xi-\hbar\omega}
\end{align}
where $\xi$, $\zeta$ label Kohn-Sham orbitals, and where the sums run over both the discrete and the continuum part of the 1-electron spectrum.

This approach notably disregards collective phenomena such as plasma oscillations, which stem from the feedback of the electrons on the potential (potential induced by the density perturbation).

In the average-atom approach, the independent-particle approximation leads to the average-atom Kubo-Greenwood formula:
\begin{align}
\text{Im}(\epsilon_\omega) =n_\text{i}\ee^2\frac{\pi}{3}& \sum_{\xi,\zeta} \left(p_\text{F}(\mu,T,\varepsilon_\xi)-p_\text{F}(\mu,T,\varepsilon_\zeta)\right)
\left| \langle\varphi_\xi \right| \vec{R}\left|\varphi_\zeta \rangle \right|^2\delta(\hbar\omega-\hbar\omega_{\zeta,\xi})
\end{align}
where $\hbar\omega_{\zeta,\xi}=\varepsilon_\zeta-\varepsilon_\xi$.
Within this approximation, the photoabsorption can be decomposed into bound-bound, bound-continuum, and continuum-continuum contributions.
Although we use here the average-atom context as an example for the discussion, a similar treatment can be performed in the case of a more detailed model yielding the contributions of the various excited states to the plasma dielectric function.

The oscillator strengths are numbers defined as follows:
\begin{align}
f_{\xi,\zeta}=\frac{2}{3}\frac{m_\text{e}}{\hbar^2}\hbar\omega_{\zeta,\xi}
\left| \langle\varphi_\xi \right| \vec{R}\left|\varphi_\zeta \rangle \right|^2
\end{align}
when $\xi$, $\zeta$ belong to the discrete part of the spectrum. In the case where either $\xi$ or $\zeta$ belongs to the continuum, this expression is to be understood as a density of oscillator strength, also called a differential oscillator strength. 

Furthermore, in the case of both $\xi$ and $\zeta$ belonging to the continuum, the dipole matrix element is a conditionally-convergent integral. In practice, in the independent-particle approximation, one can use the Ehrenfest theorem in order to recast the dipole matrix element into its acceleration form \cite{BetheSalpeterbook}:
\begin{align}
\left| \langle\varphi_\xi \right| \vec{R}\left|\varphi_\zeta \rangle \right|
=\frac{1}{m_\text{e}\omega_{\zeta,\xi}^2}
\left|\int d^3r \left\{
 \langle\varphi_\xi | \vec{r}\rangle
\vec{\nabla}_{\vec{r}}v_\text{eff}(r)
 \langle\vec{r} | \varphi_\zeta\rangle
\right\} \right|
\label{eq_accel_form}
\end{align}
where $v_\text{eff}(r)$ is the effective potential associated to the wavefunctions $\{\varphi_\xi \}$.
The role of dipole matrix element involving two continuum orbitals is rather specific to plasma physics, because it implies non-zero populations for the continuum orbitals.

\begin{figure*}[t]
\centerline{\includegraphics[width=8cm]{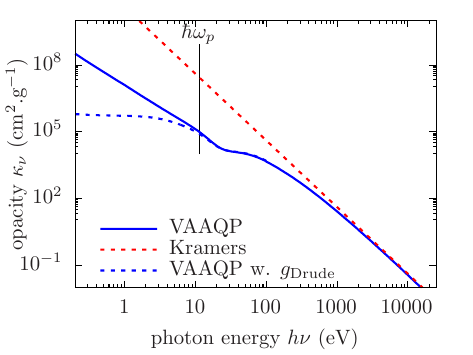}}
\caption{Continuum-continuum contribution to the opacity of a silicon plasma at 5 eV temperature and 2.36 g.cm$^{-3}$ matter denisty. Calculation using orbitals from the VAAQP model, and comparison with the Kramers formula and the opacity corrected using the $g_\text{Drude}$ function of Eq.~\eqref{eq_g_Drude}.
\label{fig_free_free}}
\end{figure*}

An example of continuum-continuum contribution to the opacity of a plasma is given in Fig.~\ref{fig_free_free}, correponding to silicon at 2.36 g.cm$^{-3}$ matter density and 5 eV temperature. In the presented calculation, the double sum, or more precisely the double integral, was performed using continuum wavefunctions obtained from the VAAQP model. Approximate methods allowing to avoid the double summation exists (see, for instance \cite{Perrot96}). One can see on this figure that, at high frequency, one recovers the Kramers classical result \cite{Kramers23,Zeldovich} involving the bare-nucleus charge, and not an effective ion charge. This is expected since the atomic response at high frequencies essentially involves the electrons located in the vicinity of the nucleus. 

\begin{figure*}[t]
\centerline{\includegraphics[width=16cm]{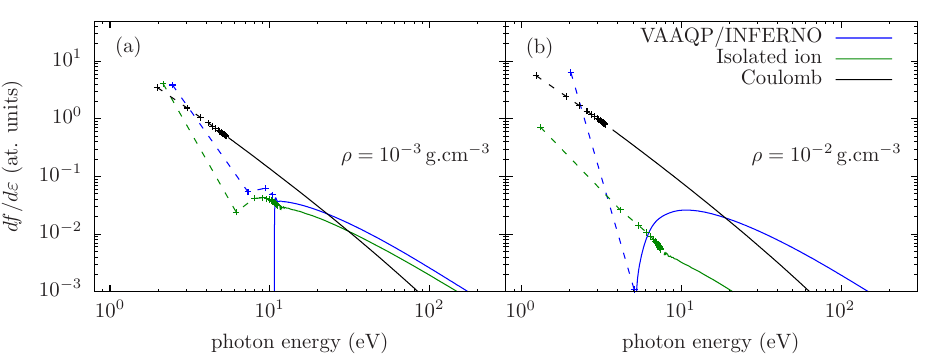}}
\caption{Differential oscillator strengths for the $4s-p$ and oscillator strengths for the $4s-np$ transitions (multiplied by the $(d\varepsilon_\xi/dn_\xi)^{-1}$ term of Eq.~\eqref{eq_continuity_osc_str}), for a silicon plasma at 5-eV temperature and matter densities of $10^{-3}$ (a) and $10^{-2}$ g.cm$^{-3}$ (b). Comparison between results from VAAQP/INFERNO (same results at these conditions), an isolated ion having the average configuration taken VAAQP/INFERNO and orbitals for a Coulomb potential with charge $Z^*$ taken from VAAQP/INFERNO. 
\label{fig_osc_strengths_drop}}
\end{figure*}

When the effective potential defining the orbitals is screened, the behavior of oscillator strengths is qualitatively modified. The underlying reasons are closely related to the limitation of the number of bound states.

For a potential having a Coulomb-tail (pure Coulomb potential of potential of an isolated-ion model), the continuity of the cross section accross the ionization threshold may easily be expressed through the matching of the two quantities:
\begin{align}
\left.f_{\xi,\zeta}\frac{1}{d\varepsilon_\xi/dn_\xi}\right|_{\varepsilon_\xi\rightarrow 0-}
=
\left.f_{\xi,\zeta}\frac{n_{\xi}^3}{Z^{*\,2}}\right|_{\varepsilon_\xi\rightarrow 0-}
=
\left.\frac{df_{\ell_\xi,\zeta}(\varepsilon)}{d\varepsilon}\right|_{\varepsilon\rightarrow 0+}
\label{eq_continuity_osc_str}
\end{align}
where $n_{\xi},\ell_{\xi}$ are the principal and orbital quantum numbers of orbital $\xi$, respectively, and where $(d\varepsilon_\xi/dn_\xi)^{-1}$ gives the density of states of the quasi-continuum of infinitely-close discrete states in the $n\rightarrow\infty$ limit.

For Coulomb-tail potential, we have a finite value of the differential oscillator strength at the threshold. With screening, the behavior of radial wavefunctions at infinity is changed: the radial wavefunctions tend to Bessel functions instead of Coulomb functions in the case of Coulomb potential. Due to the related change in the normalization coefficient, the differential oscillator strength smoothly goes  to zero \cite{Shore75}. 
This change in the behavior of oscillator strengths becomes more pronounced as density is increased, since the potential is screened over shorter distances.

Figure~\ref{fig_osc_strengths_drop} shows an illustration of this oscillator-strength drop near the photo-ionization threshold in the case of silicon at 5-eV temperature and matter densities of $10^{-3}$ and $10^{-2}$ g.cm$^{-3}$. In this figures, oscillators strengths are multiplied by the $(d\varepsilon_\xi/dn_\xi)^{-1}$ term of Eq.~\eqref{eq_continuity_osc_str}, in order to stress out the continuity with differential oscillator strengths. The results from the VAAQP model (or INFERNO, both models being in agreement) are compared to that of an isolated ion with an average configuration fixed to the VAAQP average configuration, as well as to results from a Coulomb potential with charge fixed by the VAAQP mean ionization. For both the Coulomb potential and the isolated-ion, in principle, the set of bound states is infinite, as well as the series of oscillator strengths. At the low density of $10^{-3}$ g.cm$^{-3}$, one can see that despite the qualitatively-different behavior of oscillator strengths near the photo-ionization threshold, a quantitative agreement is obtained between the VAAQP model and the isolated-ion. On the contrary, at the higher density of $10^{-2}$ g.cm$^{-3}$, the change of behavior has a larger quantitative impact.

\begin{figure}[ht]
\centerline{\includegraphics[width=8cm]{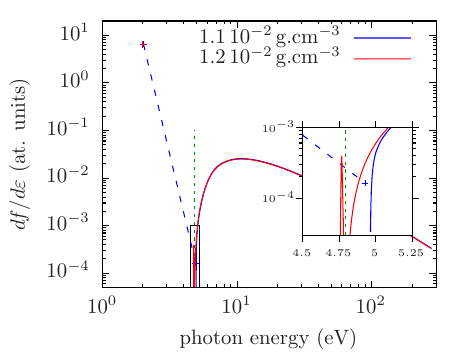}}
\caption{Differential oscillator strengths for the 4s-p and oscillator strengths for the 4s-np transitions, for a silicon plasma at 5-eV temperature and matter densities of $1.1\,10^{-2}$ and $1.2\,10^{-2}$ g.cm$^{-3}$. Results are taken from VAAQP/INFERNO (same results at these conditions). Between the two matter densities, the 5p subshell disappear and is replaced by the corresponding resonance in the $p$ continuous spectrum. The vertical dotted line in green indicates the position of the 4s photo-ionization threshold obtained from the Stewart-Pyatt formula \cite{StewartPyatt66}. 
\label{fig_osc_strengths_press_ioniz}}
\end{figure}

Using a model such as INFERNO or VAAQP, one accounts for both the decrease of the oscillator strength and the appearance of a resonance when a bound state disappears. Using an isolated ion with continuum lowering in order to suppress subshells does not account for either of these phenomena. Figure~\ref{fig_osc_strengths_press_ioniz} shows the oscillator strengths at two matter densities between which the 5p subshell gets pressure-ionized. One can easily see how the discrete oscillator strength is replaced by an equivalent contribution from a resonance in the differential oscillator strength. Thus the corresponding bound-bound channel does not disappear, but is replaced by a contribution to the bound-continuum channels. For the sake of comparison, the energy of the photoionization threshold obtained from the average-atom of isolated ion with Stewart-Pyatt IPD is also shown. The location of the threshold is in good agreement with VAAQP in this case of relatively low density, but the cross-section of an isolated-ion model would be different. Moreover, obtaining a fully continuous variation of the opacity with density requires a correct and consistent accounting for the resonances in the bound-continuum and continuum-continuum contributions to the opacity.

\subsection{Fluctuations around the average atomic state and the need for detailed modeling}
\begin{figure*}[ht]
\centerline{
\includegraphics[width=8cm]{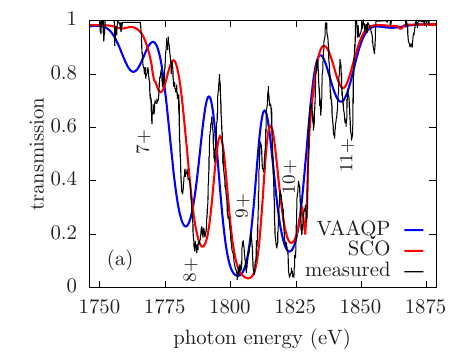}
\includegraphics[width=8cm]{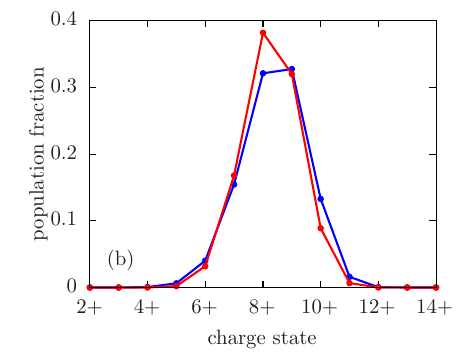}
}
\caption{Transmission of a silicon plasma from \cite{Wei08}, at areal density of 80$\mu$g.cm$^{-2}$. Estimated plasma conditions are 60 eV temperature and 45 mg.cm$^{-3}$ matter density. Comparison between results from a DCA approach based on the VAAQP model \cite{Piron13}, from the SCO approach \cite{Blenski00}, and measured transmission spectrum from \cite{Wei08}. In the calculations, an arbitrary line width of 4 eV was added to the statistical width to mimic the physical broadening and instrumental resolution.
\label{fig_Si_SG2}}
\end{figure*}

Whereas the average microstate of a whole macroscopic plasma may virtually sample many atomic excited states, the average-atom approach is based on the average \emph{atomic} state of the plasma. 

In terms of detailed atomic modeling, spectral quantities such as the opacity or emissivity may reveal the contributions of the various species which has significant populations, because different atomic states usually contribute at distinct frequencies. Even when the various spectral features are unresolved due to physical broadening, the statistical distribution among the various species yields a statistical broadening of the unresolved feature.

For that reason, detailed modeling of plasma is somehow mandatory in order to calculate realistic estimates of spectral quantities. The physical pictures underlying average-atom models can often be extended to more detailed modeling. For instance, the models \cite{Perrot87b,Blenski00} rely on various extensions of the ion-cell model to detailed configuration or superconfiguration accounting.  

The variational approach leading to the VAAQP model can be formally generalized to configurations or superconfigurations \cite{Blenski07b}. This approach was notably used to build an approximate DCA model resorting only to results from the VAAQP model (VAAQP-DCA, \cite{Piron13}).

From an average-atom standpoint, the populations of the various levels may be obtained from the analysis of fluctuations around the average atomic state \cite{Green64}. Starting from an approximate detailed model that resorts to the average-atom energies and orbitals, models of fluctuations can be used to perform a statistical approach. An exemple of such an approach is the Gaussian approximation \cite{Perrot88b}, which was also applied to the VAAQP-DCA model in \cite{Piron13}. In this context, independent particle approximation \cite{Shalitin84,Stein85,Blenski90} to the fluctuations of orbital populations leads to an overestimation of the statistical broadening, whereas correlated fluctuations \cite{Perrot88b} yields more realistic estimates. 

As an illustration of the need for detailed modeling, Fig.~\ref{fig_Si_SG2}b displays the ion charge state distributions for silicon at 60 eV temperature and 45 mg.cm$^{-3}$ matter density resulting from two detailed models: the DCA model from \cite{Piron13}, and the STA model of \cite{Blenski00}, In Fig.~\ref{fig_Si_SG2}a is shown the corresponding transimission spectra, compared to the measured spectrum from \cite{Wei08} (areal density of 80$\mu$g.cm$^{-2}$). One can easily identify the contributions from the various charge states.

Even resorting to known models, the detailed modeling of plasma still remains an implementation challenge. For elements of moderate or high atomic numbers, especially at high temperatures, the number of excited states that contribute to radiative properties may be enormous. Statistical approaches are available to reduce the number of species, leading to various levels of detail in the spectra. However, one often has to make a tradeoff between the level of detail and the completeness of the approach. Detailed modeling is also used in the collisional-radiative modeling of non equilibrium plasmas. In this context, the number of states or statistical object that may be accounted for is limited by the rank of the subsequent collisional-radiative matrix. The issue of choosing a relevant tradeoff is even more important in this context.

In the specific case of pressure ionization, the removal of some orbitals from the discrete 1-electron spectrum results in the removal of any configuration having non-zero population of these orbitals. This ultimately leads to a truncation of the charge state distribution, pushing it towards higher charge states.

\begin{figure*}[ht]
\centerline{\includegraphics[width=17cm]{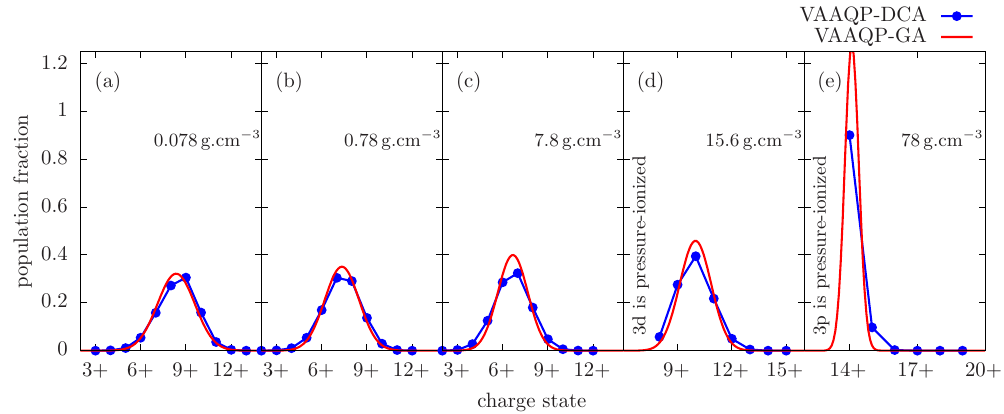}}
\caption{Charge state distributions of an iron plasma at 40-eV temperature, for various matter densities ranging from 1/100th to 10 times solid density. Comparison between a DCA approach reconstructed from the VAAQP model (VAAQP-DCA) \cite{Piron13} and an approximate treatment of this model using the gaussian approximation \cite{Perrot88b,Piron13} (VAAQP-GA).
\label{fig_CSD_Fe}}
\end{figure*}

Figure~\ref{fig_CSD_Fe} shows the charge state distributions obtained for iron at 40-eV temperature, at various matter densities. The case of 15.6 g.cm$^{-3}$ (2-fold compression) illustrate the pressure ionization of the 3d subshell, whereas to case at 78 g.cm$^{-3}$ (10-fold compression) illustrate the pressure ionization of the 3p subshell. In these cases, the results of both the reconstructed DCA model of \cite{Piron13} and an approximate, statistical treatment of the model through the gaussian approximation are displayed. One can see how the charge state distribution of the DCA model is pushed towards higher ionization stages as available configurations for the lowest ionization stages are removed. On the other hand, the gaussian approximation to fluctuations does not account properly for this cut-off but yields anyway the same qualitative trend of a narrow peak on the average charge state. 

\subsection{Collective phenomena\label{ssec_collective}}

One may expect collective phenomena to play a significant role when the frequency of the perturbing potential is lower or of the order of the plasma frequency $\omega_\text{P}$. 
\begin{align}
\omega_\text{P}=\sqrt{\frac{4\pi n_\text{e}\ee^2}{m_\text{e}}}\label{eq_plasma_freq}
\end{align}
Close to the plasma frequency, the perturbing field is resonant with the natural frequency of the free-electron gas, and its feedback may have a large impact on the plasma response. At frequencies much lower than the plasma frequency, the static collective behavior of the plasma, \textit{i.e.} the screening, prevents the electromagnetic field from propagating in the plasma, which behaves as a conductor. Indeed, below the plasma frequency, speaking in terms of dynamic conductivity rather than in terms of opacity would be more relevant.
As is clear from Eq.~\eqref{eq_plasma_freq}, collective effects impact on a frequency range which extends farther as density is increased. They thus have a specific importance for dense plasmas. 

The continuum-continuum opacity obtained from the independent-particle approximation exhibit an unphysical divergence at zero frequency, whereas the Drude model yields a finite value of the direct-current conductivity, as well as for the corresponding opacity, due to collisions. 

In \cite{Perrot96}, a very simple, heuristic approach is proposed in order to recover the Drude-like collective behavior at low frequency (very similar approaches are also described in \cite{Johnson06,Kuchiev08,Johnson09}).

From the Boltzmann equation in the relaxation-time approximation, one may derive Ziman's static conductivity \cite{Ziman61}:
\begin{align}
\gamma_\text{Ziman} = \frac{-2n_\text{e}\ee^2}{(2\pi)^3m_\text{e}^2}\frac{1}{3}\int d^3k\left\{\frac{k^2}{\omega_\text{col}(k)}  \left.\frac{\partial f_0(\varepsilon)}{\partial \varepsilon}\right|_{\varepsilon_k}\right\}
\label{eq_Ziman}
\end{align}
where $f_0$ is the free-electron energy distribution, and $\omega_\text{col}$ is the collision frequency.

In the quantum-mechanical framework, the collision frequency $\omega_\text{col}(k)$ of electrons corresponds to the net rate of elastic scattering out of the momentum $\vec{k}$. We may estimate the latter by summing the electron-ion elastic-scattering cross section, which may be estimated in the limit of weak scattering \cite{SobelmanVainshteinYukov}. One gets:
\begin{align}
\omega_\text{col}(k)=\frac{4\pi n_i}{m_\text{e} k}\sum_\ell (\ell+1) \sin^2
\left(\Delta_{k,\ell+1}-\Delta_{k,\ell}\right)
\label{eq_omega_coll}
\end{align}

The heuristical method of correction \cite{Perrot96} for the opacity is as follows. One first write the continuum-continuum photoabsorption cross-section in the low-frequency limit in terms of the scattering cross-section (the method was given in \cite{Somerville64,Ashkin66}). Then by analogy with Ziman's formula Eq.~\eqref{eq_Ziman} resorting to the collision frequency of Eq.~\eqref{eq_omega_coll}, one identifies the correcting factor which allows one to recover Ziman's result. 
\begin{align}
g_\text{Drude}(k,\omega)=\frac{\omega^2}{\omega^2+\omega_\text{col}^2(k)}
\label{eq_g_Drude}
\end{align}
This factor introduces a Drude-like behavior in the low-frequency part of the spectral opacity. 
Figure~\ref{fig_free_free} shows the effect of the correcting function $g_\text{Drude}$ in the case of silicon at 5 eV temperature and 2.36 g.cm$^{-3}$ matter denisty. As may be seen from the figure, this correction has a strong impact below the plasma frequency.

At low frequencies, the complex refraction index may also have a significant imaginary part. The assumption $n_\omega^\text{ref}=1$, often used in the dielectric regime ($\omega >> \omega_\text{P}$), is no more valid and a more realistic estimate is required. In \cite{Perrot96}, a simple estimate obtained from the Drude formula is used. However, since $\text{Im}(\epsilon_\omega)$ is known, one can also use the Kramers-Kronig relations to obtain $\text{Re}(\epsilon_\omega)$:
\begin{align}
\text{Re}(\epsilon_\omega)
=\epsilon_0+
\frac{1}{\pi}\mathcal{PP}\int_{-\infty}^{+\infty}
d\omega'\left\{
\frac{\text{Im}(\epsilon_\omega)}
{\omega'-\omega}
\right\}
\end{align}
Then, one uses Eqs~\eqref{eq_real_nref} and \eqref{eq_sigma_abs} to obtain the opacity.

\begin{figure}[ht]
\centerline{\includegraphics[width=8cm]{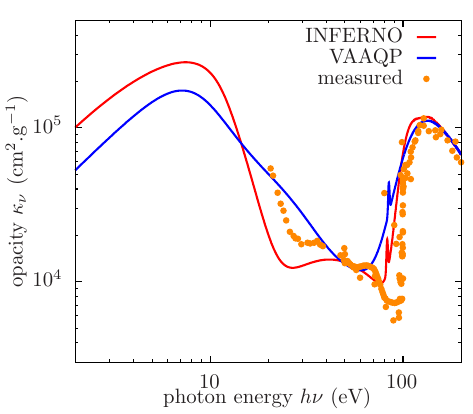}}
\caption{Opacity of cold silicon at solid density (2.36 g.cm$^{-3}$) in the visible-to-XUV range, typical of the L and M edges. Comparison between results from INFERNO and VAAQP at a temperature of 2.5 eV, using the heuristical accounting for collective effects, and measurments of the opacity of cold silicon \cite{Henkereport}.  
\label{fig_si_cold_opacity}}
\end{figure}

Figure~\ref{fig_si_cold_opacity} displays the results of the present approach \cite{Piron18}, using the heuristical coefficient $g_\text{Drude}$ and the refraction index obtained from the Kramers-Kronig relation, on the case of silicon at solid density and 2.5 eV temperature, using the VAAQP and INFERNO models. A rather good agreement is found with measurments performed on cold solid silicon \cite{Henkereport}. In fact, cold solid silicon being a metal, it is not so surprising that plasma models  can give reasonable agreement at low temperature with measurments performed on metallic silicon.

\subsection{Self-consistent linear response}
Using a time-dependent density functional formalism \cite{Stott80b,Zangwill80,Runge84,Dhara87,Gosh88} to calculate the linear response leads to a self-consistent scheme for the frequency-dependent density perturbation and induced potential:

\begin{align}
&\delta n_\omega(\vec{r})
=
\int d^3r'\left\{\mathcal{D}_\omega^{R}(\vec{r},\vec{r}')
\delta v_{\text{ext},\omega}(\vec{r}')\right\}
=
\int d^3r'\left\{\mathcal{D}_{0,\omega}^{R}(\vec{r},\vec{r}')
\left(\delta v_{\text{ext},\omega}(\vec{r}')+\delta v_{\text{ind},\omega}(\vec{r}')\right)\right\}
\\
&\delta v_{\text{ind},\omega}(\vec{r})
=\frac{\ee^2}{4\pi\varepsilon_0}
\int d^3r'\left\{
\frac{\delta n_\omega(\vec{r}')}{|\vec{r}-\vec{r}'|}\right\}
+\left.\frac{\partial \mu_\text{xc}(n)}{\partial n}\right|_{n(\vec{r})} \delta n_\omega(\vec{r})
\end{align}
where we limit ourselves to the adiabatic local density approximation to the exchange-correlation.

\begin{figure}[ht]
\centerline{\includegraphics[width=8cm]{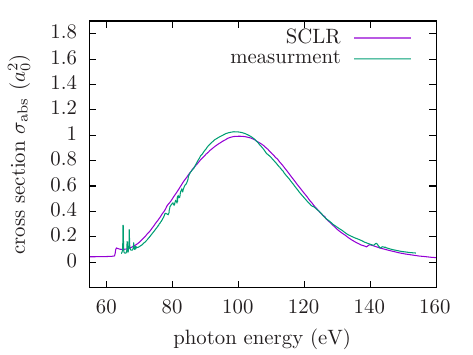}}
\caption{Photoabsorption cross section of neutral Xe in the XUV region typical of the N-edge. Comparison between the self-consistent linear response of the DFT atom, using Gunarsson-Lundqvist exchange-correlation term (same model as in \cite{Zangwill80}), the independent-particle approximation and measurments from \cite{Haensel69}.
\label{fig_LR_xe}}
\end{figure}

These equations are solved in \cite{Zangwill80} to obtain photoabsorption cross-sections of neutral rare gases, which do not involve continuum-continuum channels. However, in these calculations, the accounting for channel mixing between bound-bound and bound-continuum contributions has significant impact on the photoabsorption cross-section, especially near the photo-ionization edge. Figure~\ref{fig_LR_xe} displays the result of a self-consistent dynamic linear response calculation using the same model as \cite{Zangwill80}, on one of their case of application. Results are in close agreement both with those of \cite{Zangwill80} and with the measurement of \cite{Haensel69} on liquid xenon, and exhibit significant impact of channel mixing.

In the case of a plasma, the contribution of continuum-continuum transitions cause difficulties, since they involve non-localized wavefunctions. The cluster expansion, on which the VAAQP model is based, is an efficient method to substract the non-integrable contribution of the homogeneous medium. From the corresponding self-consistent linear-response formalism, a sum rule for the atomic dipole was derived in \cite{Blenski06}:
\begin{align}
\int d^3r\left\{z\delta n_\omega(\vec{r})\right\}
=\frac{1}{m_e\omega^2\left(1-\frac{\omega_\text{P}^2}{\omega^2}\right)}
\left(
-\int  d^3r \left\{
\delta n_\omega(\vec{r})
\frac{\partial v_\text{trial}(\vec{r})}{\partial z}
\right\}
+\int  d^3r \left\{
\delta v_{\text{ind},\omega}(\vec{r})
\frac{\partial n(\vec{r})}{\partial z}
\right\}
\right)
\label{eq_sum_ehrenfest}
\end{align}
Formally, this relation plays the same role as the switching between length and acceleration form of the dipole matrix element, but in the context of the self-consistent linear response. More precisely, the factor in front of the right-hand side stems from the contribution of the homogeneous medium and causes a singularity at the plasma frequency. This is due to the accouting for the induced field in the homogeneous plasma response, while disregarding collisions, which would saturate this resonant behavior. If one take the factor in the $\omega >> \omega_\text{P}$ limit, then the first term of Eq.~\eqref{eq_sum_ehrenfest} right-hand side may be recovered from Eq.~\eqref{eq_accel_form}. The second term is purely due to the accounting for induced potential, that is, for self-consistency in the dynamic behavior of the displaced-electron density. These effects potentially include the coupling with plasmons or the mixing among the various particle-hole channels.

Practical use of Eq.~\eqref{eq_sum_ehrenfest} was successfully implemented in the TF model \cite{Blenski13,Caizergues14}. The TF model is a relevant first-step, because the VAAQP model is rigorously equivalent to the usual TF model when the electron density is taken in the TF approximation \cite{PironPhD}. However, the TF model leads to unphysical results for the radiative properties, especially at high frequencies \cite{Ishikawa98b}.

In the quantum version of the VAAQP model, although progress was achieved in the understanding of the problem, application of the self-consistent linear-response approach still lead to unconclusive results \cite{Caizergues16}. In particular, direct check of Eq.~\eqref{eq_sum_ehrenfest} from an implementation of the self-consistent linear response using standard methods described in \cite{Zangwill80,MahanSubbaswamy} failed. Self-consistent linear response in the context of dense plasmas thus remains an open problem, at least from an implementation standpoint.

\section{Some words on collisional processes}
\begin{figure}[ht]
\centerline{
\includegraphics[width=8cm]{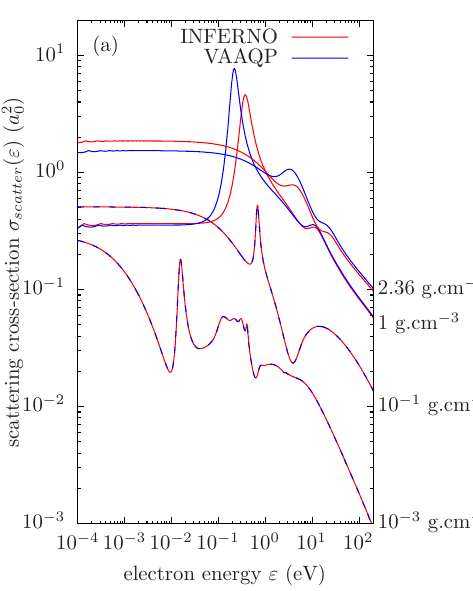}
\hspace{1cm}
\includegraphics[width=8cm]{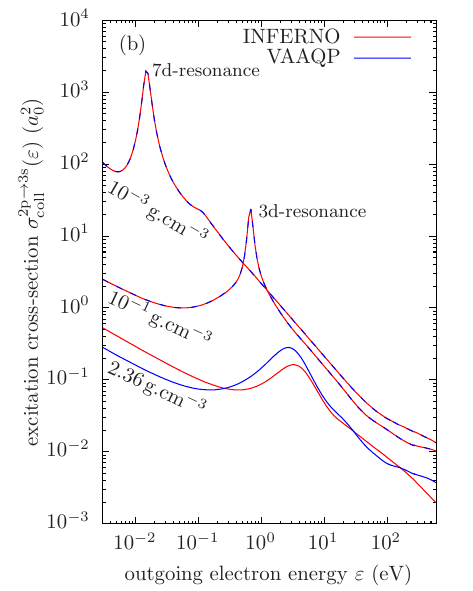}
}
\caption{ Total electron-ion elastic-scattering cross section (a) and 2p$\rightarrow$3s collisional-excitation cross section (b) for a silicon plasma at a temperature of 5 eV, for various values of the matter density. Comparison between results from the INFERNO and VAAQP models, which are in close agreement up to matter density of 0.1 g.cm$^{-3}$.
\label{fig_electron_scatter}
\label{fig_coll_excitation}
}
\end{figure}

In the heuristical bridging to Ziman formula (see Sec.\ref{ssec_collective}), we obtained the collisional frequency considering the net rate of electron-ion elastic-scattering out of momentum $\vec{k}$. In this context, we used the electron-ion elastic-scattering cross-section stemming from the limit of weak collisions. The latter only depends on the wavefunctions through the phase shifts (see, for instance \cite{SobelmanVainshteinYukov}):
\begin{align}
\sigma_{\text{scatter}}(\varepsilon_k)
=a_0^2\,\frac{4\pi}{k^2}
\sum_{\ell}(2\ell+1)\sin^2(\Delta_{k,\ell})
\end{align}
Elastic scattering of electrons by ions may be categorized as an elementary collisional atomic process, even if it does not change the ion electronic state. Figure~\ref{fig_electron_scatter}a displays the total electron-ion elastic-scattering cross section for a silicon plasma at a temperature of 5 eV, for various values of the matter density. Below matter density of 1 g.cm$^{-3}$, results from the INFERNO and VAAQP models agree well .

The distorted wave approximation (see, for instance \cite{SobelmanVainshteinYukov}) is widely used for the calculation of cross sections of collisional processes for isolated ions \cite{Sampson09,Peyrusse99}. A straightforward approach consists in extending heuristically this method to models of pressure-ionized plasmas. For example one may calculate the collisional-excitation cross section by generalizing the configuration-averaged collision strength \cite{Peyrusse99} to the orbitals and fractional occupation numbers given by an average-atom model such as INFERNO or VAAQP.

As an illustration, we display in Fig.~\ref{fig_coll_excitation}b the 2p$\rightarrow$3s collisional-excitation cross section, obtained using the distorted-waves approach using the INFERNO and VAAQP models, respectively. The chosen case is a silicon plasma at 5-eV temperature, for matter densities of $10^{-3}$, $10^{-1}$, and 2.36 g.cm$^{-3}$. At high energies, the Born behavior (plane-wave incoming and outgoing electrons) is recovered. At matter density of $10^{-3}$ g.cm$^{-3}$, the 7d bound orbital is delocalized, but is still present as a sharp resonance in the continuum.  At matter density of $10^{-1}$ g.cm$^{-3}$, the same occurs for the 3d bound orbital.

As a direct consequence of these resonances, we obtain sharp, quasi-discrete features in the corresponding cross-sections. These near-threshold sharp features may be seen as the remnants of the correponding dielectronic recombination channels, namely: 2p$\rightarrow$3s,7d, and 2p$\rightarrow$3s,3d.

One thus can see how, prior to effectively removing them, screening of the potential may redispatch the collisional channels. Part of the dielectronic recombination channels then becomes collisional excitation.

However, one may question the general approach to collisional process in the context of dense-plasma models. In fact, collisional processes are  introduced as a perturbative accounting for continuum electrons, supplementing an ion Hamiltonian that disregards them (see, for instance \cite{SobelmanVainshteinYukov}). Yet, models of pressure-ionized plasma account for continuum electron as part of the ion electronic structure. In this framework, one should probably adapt, re-interpret or redefine the whole approach to collisional processes. To the best of the author's knowledge, such a rigorous approach to collisional processes in pressure-ionized plasma is an open question.

%

\section{Conclusion}
Atomic models of dense plasmas in themselves are still an active field of research, facing open questions as regards their theoretical founding. The main shortfalls of models based on the notion of continuum lowering are rather well identified. However, models used in order to go beyond the continuum lowering picture are still mostly based on the picture of a Wigner-Seitz cavity. This picture may be seen as a practical, heuristical way of introducing the pressure ionization in the models but seems poorly motivated, especially for weakly and strongly coupled plasmas. 

Progress are still ongoing towards a better modeling of pressure ionization, closer to the first-principles. That means obtaining pressure ionization directly as a consequence of the plasma structure. Recent research efforts were done in this direction with the models based on the QHNC approach and the VAMPIRES model. These models may be viewed as steps in the understanding of the problem but for sure, they do not exhaust the theoretical challenge of the consistent modeling of  nucleus-electron plasmas.

It is to be expected that defining a relevant notion of ion may not be possible at all plasma conditions. Defining the validity domain of atomic physics of plasma, if not of a particular model, remains among the most challenging issues.

Finding a satisfactory model of ions in dense plasmas is the essential first step towards applying many methods and notions from atomic physics. This notably includes the calculation of radiative properties or atomic processes typically involved in collisional-radiative modeling.

As regards radiative properties, most approaches are based on the independent-particle approximation. In these methods, the high number of excited states in high-temperature, mid-to-high-Z plasmas may still constitute an implementation challenge, requiring a tradeoff between completeness and level of detail.

Self-consistent linear-response was successfully applied as regards bound electrons and their related contribution to radiative properties since the 80's \cite{Zangwill80,Zangwill84,Doolen87}. However, despite a significant theoretical effort \cite{Blenski06,Caizergues14,Caizergues16}, the consistent treatment of the continuum electron remains an issue in the quantum framework. Yet, this would constitute a important step towards a first-principles approach to collective effects on the radiative properties. Heuristical approaches to collective effects are known \cite{Perrot96,Johnson06,Kuchiev08,Johnson09}. However, all of them are in fact very similar and a better-founded approach would be of great interest.

Addressing collisional processes is especially required for the collisional-radiative modeling of dense plasmas. Work is in progress to study the collisional processes in the framework of fully quantum models of screening in dense plasmas. From a direct application of the distorted-waves approach, it appears that screening of the potential can results in a different dispatching of transitions among excitation, ionization and dielectronic channels. However, the theoretical justification for applying the distorted-wave approach to models that accounts for continuum electrons deserves deeper investigation. 

Some dense-plasma models, such as INFERNO, have now been studied and used for many years but experimental checks of their validity are scarce and do not really allow to discriminate among the various models. Equation of state measurments often have large experimental uncertainties and rarely access the temperature of the plasma. On the other hand, measurments of radiative properties are most often performed on diluted plasma (see for instance \cite{Davidson88,Bruneau90,DaSilva92,Bailey07,Bailey15}), addressing regimes in which differences among models are not pronounced. Direct-current conductivity measurments \cite{Renaudin02,Korobenko07} or X-ray Thomson scattering may adress relevant regimes but usually requires to make a step further in the modeling in order to interpret the measurments \cite{Gregori04,Sperling17}.

Efforts to improve atomic models of dense plasma are timely, in view of the growing concern for understanding the warm-dense matter, with applications to stellar astrophysics and planetology in mind. These efforts are also in pace with the recent progresses of experiments on warm and hot dense plasmas, impulsed by the advent of new facilities and experimental platforms. One may cite for instance the recent convergent-spherical shockwave experiments at NIF \cite{Swift18}, which gives access to equation-of-state data at Gbar pressures, the opacity measurments of compressed plasma at OMEGA \cite{Hu22}, the measurment of spectral emission of dense, near-equilibrium plasma using buried layers at ORION \cite{Hoarty13}, or the experiments on the photoionization of metals using tunable X-ray free-electron laser at LCLS \cite{Ciricosta12}. These recent progresses of experimental techniques may allow to better question the models limitations.

\section*{Acknowledgements}
The author would like to thank T. Blenski, with whom he has been working on the present subject for many years. The author's present understanding of the field owes much to shared work and discussions with him.  

The author would also like to thank F. Rosmej, guest editor of the present special issue, for giving him this opportunity to share his view on the subject. 

The author would like to thank B. Cichocki, who was involved in many studies reviewed here, including the early stage of development of the VAMPIRES model, and C. Caizergues, who worked on the linear response of the VAAQP model as a PhD student, co-supervised by T. Blenski and the author.

Part of the material presented in this paper was initially prepared for an invited talk given at the 20th conference on "Atomic Processes in Plasmas (APIP)", in Gaithersburg, April 12th 2019. Another part was prepared for an invited talk given at the workshop ``Astrophysics with High Power Lasers and Laboratory Plasmas'', in Paris, December 16th, 2021. The author would like to thank their respective organizers Y. Ralchenko and A. Ciardi for giving him the occasion to gather these pieces of work that would otherwise have been scattered. The author also aknowledges a short but enlightening discussion with O. Peyrusse on the occasion of the 2019 APIP conference.


\bibliographystyle{unsrt}
\bibliography{/home/piron/biblio/biblio-utf8}

\end{document}